\documentclass[
    twocolumn,
	prl,
	preprintnumbers,superscriptaddress,
	nofootinbib]{revtex4-2}
	  
\pdfoutput=1

\usepackage{graphicx}
\usepackage{enumitem}
\usepackage{latexsym}
\usepackage{amsfonts}
\usepackage{amssymb}
\usepackage{xcolor}
\usepackage[export]{adjustbox}
\usepackage{amsmath}
\usepackage[thinlines]{easytable}
\usepackage{verbatim}
\usepackage{bm}
\usepackage{float}
\usepackage{multirow}
\usepackage{xspace}
\usepackage[normalem]{ulem}
\usepackage[
pdfauthor={Jeremy Sakstein}]{hyperref}
\usepackage{tabularx}

\renewcommand{\tilde}{\widetilde} 
\newcommand{\beq}{\begin{equation}}
\newcommand{\eeq}{\end{equation}}
\newcommand{\bea}{\begin{eqnarray}}
\newcommand{\eea}{\end{eqnarray}}

\newcommand{\refjnl}[1]{{\rm#1}}

\def\aj{\refjnl{AJ}}                   % Astronomical Journal
\def\apj{\refjnl{ApJ}}                 % Astrophysical Journal
\def\apjl{\refjnl{ApJ}}                % Astrophysical Journal, Letters
\def\apjs{\refjnl{ApJS}}               % Astrophysical Journal, Supplement
\def\aap{\refjnl{A\&A}}                % Astronomy and Astrophysics
\def\mnras{\refjnl{MNRAS}}             % Monthly Notices of the RAS
\def\ssr{\refjnl{Space~Sci.~Rev.}}     % Space Science Reviews
\usepackage{pdfbase}[2017/03/16]
\usepackage{xparse,ocgbase}
\usepackage{xcolor,calc}
\usepackage{tikzpagenodes,linegoal}
\usetikzlibrary{calc}
\usepackage{tcolorbox}

\ExplSyntaxOn
\let\tpPdfLink\pbs_pdflink:nn
\let\tpPdfAnnot\pbs_pdfannot:nnnn\let\tpPdfLastAnn\pbs_pdflastann:
\let\tpAppendToFields\pbs_appendtofields:n
\def\tpPdfXform{\pbs_pdfxform:nnnnn{1}{1}{}{}}
\let\tpPdfLastXform\pbs_pdflastxform:
\let\cListSet\clist_set:Nn\let\cListItem\clist_item:Nn
\ExplSyntaxOff

\usepackage{pdfbase}[2017/03/16]
\usepackage{xparse,ocgbase}
\usepackage{xcolor,calc}
\usepackage{tikzpagenodes,linegoal}
\usetikzlibrary{calc}
\usepackage{tcolorbox}

\ExplSyntaxOn
\let\tpPdfLink\pbs_pdflink:nn
\let\tpPdfAnnot\pbs_pdfannot:nnnn\let\tpPdfLastAnn\pbs_pdflastann:
\let\tpAppendToFields\pbs_appendtofields:n
\def\tpPdfXform{\pbs_pdfxform:nnnnn{1}{1}{}{}}
\let\tpPdfLastXform\pbs_pdflastxform:
\let\cListSet\clist_set:Nn\let\cListItem\clist_item:Nn
\ExplSyntaxOff

\makeatletter
\NewDocumentCommand{\tooltip}{%
  ssssO{\ifdefined\@linkcolor\@linkcolor\else blue\fi}mO{yellow!20}mO{0pt,0pt}%
}{{%
  \leavevmode%
  \IfBooleanT{#2}{%
    \ocgbase@new@ocg{tipOCG.\thetcnt}{%
      /Print<</PrintState/OFF>>/Export<</ExportState/OFF>>%
    }{false}%
    \xdef\tpTipOcg{\ocgbase@last@ocg}%
    \ocgbase@add@ocg@to@radiobtn@grp{tool@tips}{\ocgbase@last@ocg}%
  }%
  \tpPdfLink{%
    \IfBooleanTF{#4}{%
      /Subtype/Link/Border[0 0 0]/A <</S/SetOCGState/State [/Toggle \tpTipOcg]>>
    }{%
      /Subtype/Screen%
      /AA<<%
        \IfBooleanTF{#3}{%
          /E<</S/SetOCGState/State [/Toggle \tpTipOcg]>>%
        }{%
          \IfBooleanTF{#2}{%
            /E<</S/SetOCGState/State [/ON \tpTipOcg]>>%
            /X<</S/SetOCGState/State [/OFF \tpTipOcg]>>%
          }{
            \IfBooleanTF{#1}{%
              /E<</S/JavaScript/JS(%
                var fd=this.getField('tip.\thetcnt');%
                if(typeof(click\thetcnt)=='undefined'){%
                  var click\thetcnt=false;%
                  var fdor\thetcnt=fd.rect;var dragging\thetcnt=false;%
                }%
                if(fd.display==display.hidden){%
                  fd.delay=true;fd.display=display.visible;fd.delay=false;%
                }else{%
                  if(!click\thetcnt&&!dragging\thetcnt){fd.display=display.hidden;}%
                  if(!dragging\thetcnt){click\thetcnt=false;}%
                }%
                this.dirty=false;%
              )>>%
            }{%
              /E<</S/JavaScript/JS(%
                var fd=this.getField('tip.\thetcnt');%
                if(typeof(click\thetcnt)=='undefined'){%
                  var click\thetcnt=false;%
                  var fdor\thetcnt=fd.rect;var dragging\thetcnt=false;%
                }%
                if(fd.display==display.hidden){%
                  fd.delay=true;fd.display=display.visible;fd.delay=false;%
                }%
               this.dirty=false;%
              )>>%
              /X<</S/JavaScript/JS(%
                if(!click\thetcnt&&!dragging\thetcnt){fd.display=display.hidden;}%
                if(!dragging\thetcnt){click\thetcnt=false;}%
                this.dirty=false;%
              )>>%
            }%
            /U<</S/JavaScript/JS(click\thetcnt=true;this.dirty=false;)>>%
            /PC<</S/JavaScript/JS (%
              var fd=this.getField('tip.\thetcnt');%
              try{fd.rect=fdor\thetcnt;}catch(e){}%
              fd.display=display.hidden;this.dirty=false;%
            )>>%
            /PO<</S/JavaScript/JS(this.dirty=false;)>>%
          }%
        }%
      >>%
    }%
  }{{\color{#5}#6}}%
  \sbox\tiptext{%
    \IfBooleanT{#2}{%
      \ocgbase@oc@bdc{\tpTipOcg}\ocgbase@open@stack@push{\tpTipOcg}}%
    \tcbox[colframe=black,colback=#7,size=fbox,arc=1ex,sharp corners=southwest]{#8}%
    \IfBooleanT{#2}{\ocgbase@oc@emc\ocgbase@open@stack@pop\tpNull}%
  }%
  \cListSet\tpOffsets{#9}%
  \edef\twd{\the\wd\tiptext}%
  \edef\tht{\the\ht\tiptext}%
  \edef\tdp{\the\dp\tiptext}%
  \tipshift=0pt%
  \IfBooleanTF{#2}{%
    \setlength\whatsleft{\linegoal}%
  }{%
    \measureremainder{\whatsleft}%
  }%
  \ifdim\whatsleft<\dimexpr\twd+\cListItem\tpOffsets{1}\relax%
    \setlength\tipshift{\whatsleft-\twd-\cListItem\tpOffsets{1}}\fi%
  \IfBooleanF{#2}{\tpPdfXform{\tiptext}}%
  \raisebox{\heightof{#6}+\tdp+\cListItem\tpOffsets{2}}[0pt][0pt]{%
    \makebox[0pt][l]{\hspace{\dimexpr\tipshift+\cListItem\tpOffsets{1}\relax}%
    \IfBooleanTF{#2}{\usebox{\tiptext}}{%
      \tpPdfAnnot{\twd}{\tht}{\tdp}{%
        /Subtype/Widget/FT/Btn/T (tip.\thetcnt)%
        /AP<</N \tpPdfLastXform>>%
        /MK<</TP 1/I \tpPdfLastXform/IF<</S/A/FB true/A [0.0 0.0]>>>>%
        /Ff 65536/F 3%
        /AA <<%
          /U <<%
            /S/JavaScript/JS(%
              var fd=event.target;%
              var mX=this.mouseX;var mY=this.mouseY;%
              var drag=function(){%
                var nX=this.mouseX;var nY=this.mouseY;%
                var dX=nX-mX;var dY=nY-mY;%
                var fdr=fd.rect;%
                fdr[0]+=dX;fdr[1]+=dY;fdr[2]+=dX;fdr[3]+=dY;%
                fd.rect=fdr;mX=nX;mY=nY;%
              };%
              if(!dragging\thetcnt){%
                dragging\thetcnt=true;Int=app.setInterval("drag()",1);%
              }%
              else{app.clearInterval(Int);dragging\thetcnt=false;}%
              this.dirty=false;%
            )%
          >>%
        >>%
      }%
      \tpAppendToFields{\tpPdfLastAnn}%
    }%
  }}%
  \stepcounter{tcnt}%
}}
\makeatother
\newsavebox\tiptext\newcounter{tcnt}
\newlength{\whatsleft}\newlength{\tipshift}
\newcommand{\measureremainder}[1]{%
  \begin{tikzpicture}[overlay,remember picture]
    \path let \p0 = (0,0), \p1 = (current page.east) in
      [/utils/exec={\pgfmathsetlength#1{\x1-\x0}\global#1=#1}];
  \end{tikzpicture}%
}

\newcommand{\dm}{_{\rm DM}}
\newcommand{\vF}{\vec{\mathcal{F}}}
\newcommand{\vx}{\vec{\xi}}
\newcommand{\Qq}{ \mathcal{Q}_q(t)}

\DeclareRobustCommand{\okina}{%
  \raisebox{\dimexpr\fontcharht\font`A-\height}{%
    \scalebox{0.8}{`}%
  }%
}

\newcommand{\dd}{\mathrm{d}}

\renewcommand{\vec}{\bm}

\allowdisplaybreaks
\begin{document}

\title{Dark Matter-Induced Stellar Oscillations}
\author{Jeremy Sakstein} \email{sakstein@hawaii.edu}
\affiliation{Department of Physics \& Astronomy, University of Hawai\okina i, Watanabe Hall, 2505 Correa Road, Honolulu, HI, 96822, USA}
\author{Ippocratis D.~Saltas} \email{saltas@fzu.cz}
\affiliation{CEICO, Institute of Physics of the Czech Academy of Sciences, Na Slovance 2, 182 21 Praha 8, Czechia}

\date{\today}

\begin{abstract}
It has been hypothesized that dark matter is comprised of ultra-light bosons whose collective phenomena can be described as a scalar field undergoing coherent oscillations.~Examples include axion and fuzzy dark matter models.~In this ultra-light dark matter scenario, the harmonic variation in the field's energy-momentum tensor sources an oscillating component of the gravitational potential that we show can resonantly-excite stellar oscillations.~A mathematical framework for predicting the amplitude of these oscillations is developed, which reveals that ultra-light dark matter predominantly excites p-modes of degree $l=1$.~An  investigation of resonantly-excited solar oscillations is presented, from which we conclude that dark matter-induced oscillations of the Sun are likely undetectable.~We discuss  prospects for constraining ultra-light dark matter using other stellar objects.
\end{abstract}

\maketitle

Ultra-light dark matter (ULDM) is an interesting alternative to the cold dark matter (CDM) paradigm \cite{Marsh:2015xka,Hui:2016ltb,Ferreira:2020fam,Hui:2021tkt,Chadha-Day:2021szb}.~In this scenario, dark matter is comprised of light bosons with a large occupancy number whose collective phenomena is best-described by classical scalar waves.~In this work we focus on scalar (spin-zero) ULDM models such as axions and fuzzy dark matter.~ULDM is a tantalizing theory of dark matter (DM) because it can explain several of the small-scale astrophysical anomalies such as the cusp vs.~core and missing satellite problems that CDM cannot account for without appealing to highly-uncertain baryonic effects \cite{Hu:2000ke, Hui:2016ltb,Ferreira:2020fam}.~Theoretically, ULDM particles are motivated by the QCD axion (see e.g., \cite{Peccei:1977hh,Weinberg:1977ma,Wilczek:1977pj,Zhitnitsky:1980tq}), the plethora of axion-like particles predicted by string theory (see e.g., \cite{Svrcek:2006yi,Arvanitaki:2009fg}), and superfluid DM \cite{Berezhiani:2015bqa,Khoury:2021tvy}.~

In this paper, we show that the collective wave-like behavior of ULDM can resonantly-excite stellar oscillations with frequencies close to the DM particle's mass.~The harmonic variation of the scalar's energy-momentum tensor induces an oscillating component of the Newtonian potential\footnote{ULDM-induced oscillations in the metric potentials have previously been tested using pulsar timing arrays \cite{Khmelnitsky:2013lxt}, laser interferometers \cite{Aoki:2016kwl}, and binary pulsars \cite{Blas:2016ddr,Blas:2019hxz}.} \cite{Khmelnitsky:2013lxt,Aoki:2016kwl,Blas:2016ddr} that acts as a driving term for stellar oscillations.~We derive a mathematical framework for calculating the amplitude of DM-induced stellar oscillations, and find that p-modes ($l=1$) are predominantly excited.~We then apply this formalism to a model for the Sun and find that DM-induced solar oscillations are likely undetectable.~We therefore conclude by discussing the prospects for testing ULDM with other stellar objects.

The action for the ULDM scalar $\phi$ coupled to gravity is
\begin{equation}
    \label{eq:action}
    S = \int\dd^4x\sqrt{-g}\left[\frac{R}{16\pi G} -  \frac12\nabla_\mu\phi\nabla^\mu\phi-\frac12m^2\phi^2\right].
\end{equation}
In the rest frame of the DM halo (where the ULDM has zero velocity) the equation of motion for $\phi$, $(\Box-m^2)\phi=0$, has solution $    \phi=\phi_0\cos( mt)$, where $\phi_0$ is a constant and we have chosen appropriate initial conditions.\footnote{We neglect  spatial variation of the field since the scale we are interested in ($\sim$$R_\odot$) is far smaller than the scale over which the DM density varies (the DM halo scale radius).}~Under this assumption, the energy-momentum tensor is $T_{\mu\nu}=\mathrm{diag}(\rho\dm, P\dm, P\dm , P\dm)$, 
% \begin{equation}
% \label{eq:EMtensor_rest_frame}
%     T_{\mu\nu}=\begin{pmatrix}
%   \rho & 0  \\
%   0 & P\delta_{ij}  \\
%  \end{pmatrix},
% \end{equation}
where the density and pressure are $\rho\dm=\frac12m^2\phi_0^2$ and $P\dm=-\rho\dm\cos(\omega t)$  with $\omega=2m$ \cite{Khmelnitsky:2013lxt,Aoki:2016kwl, Hui:2016ltb, Hui:2021tkt}.~Thus, in the galactic rest frame, ULDM has a constant density\footnote{The wavelength of the DM particles we consider in this work is many-orders-of-magnitude larger than a solar radius, implying that the accumulation of dark matter inside the star is negligible.} and an oscillating pressure.~%\sout{~The oscillating pressure causes oscillations in the spatial part of the metric $g_{ij}$ but not the Newtonian potential \cite{Khmelnitsky:2013lxt}.~} \is{I still don't agree with this statement..~There is time dependence to both potentials due to the axionic resonance, so cannot see how we can state that it is only the relativistic one that picks up time dependence.}\js{Move this later on.}

A star moving with velocity $\vec{v}^\star$ through the DM halo will experience a Lorentz-transformed energy-momentum tensor given by \cite{Aoki:2016kwl}
$T_{00}=\rho\dm\gamma^2\left[1-{v^\star}^2\cos(\omega t')\right]$, $T_{0i}=\rho\dm\gamma^2v^\star_i\left[1-\cos(\omega t')\right]$, and $ T_{ij}= -\rho\dm\cos(\omega t')+\rho\dm\gamma^2 v^\star_i v^\star_j\left[1-\cos(\omega t')\right]$, 
where $t'=\gamma\left(t+\vec{v^\star}\cdot\vec{x}\right)$.
In the star's rest frame, the DM density oscillates, which gives rise to an oscillating Newtonian potential.~Expanding the metric as $\dd s^2=-(1+2\Phi)\dd t^2+\left(1+2\Psi\right)\delta_{ij}\dd x^i\dd x^j$, the Newtonian potential $\Phi$ in this frame is (we will not need $\Psi$ in what follows) \footnote{Note that there is an oscillating Newtonian potential in the limit $v^\star\rightarrow0$ but this is a pure gauge-mode that does not impact the dynamics.~Only objects in motion with respect to the DM rest frame feel an oscillating gravitational force.} 
\begin{align}
\label{eq:oscillating_potentials}
    \Phi(\vec{x},t)=-\frac{\pi G\rho\dm}{m^2}\cos(\omega t').
    %\quad\textrm{and}\quad
    %\Psi(\vec{x},t)=\frac{\pi G\rho\dm}{m^2}\cos(\omega t').
\end{align}
The oscillating Newtonian potential 
%$\Phi$ 
acts as a new gravitational source in the stellar structure equations, and, as we will see momentarily, can excite the star's oscillatory modes.~It should be understood as a perturbative correction to the static gravitational potential $\Phi_0(\vec{x},t)$ sourced by baryonic matter as $\Phi^{\rm (total)} = \Phi_0(\vec{x},t) + \Phi(\vec{x},t)$.~ 

%two scalar gravitational potentials in this frame are

Our starting point for deriving the effects of the oscillating Newtonian potential upon stellar oscillations is the mathematical formalism for describing linear, adiabatic perturbations of the star about its spherically-symmetric equilibrium configuration \cite{1980tsp..book.....C,1989nos..book.....U,2001A&A...370..136S,2001A&A...373..916L,2005MNRAS.360..859C,aerts2010asteroseismology,Lopes:2014dba,Lopes:2015pca}.~In this formalism, the fundamental variable of interest is the  Lagrangian displacement $\vec{\xi}=\vec{\delta r}$ (i.e., the position of a fluid element is $\vec{r}=r_0\hat{\vec{r}}+\vec{\delta r}$ with $r_0$ the time-independent equilibrium position), which satisfies $|\vec{\xi}|/R\ll1$ where $R$ is the star's radius.~Displaced fluid elements result in (linear) perturbations to the equilibrium quantities.~In what follows we will work with Eulerian fluid variables and  will denote equilibrium quantities with subscript zeros and linear perturbations with tildes e.g., the linearly perturbed Eulerian density is $\rho(\vec{x},t)=\rho_0(r_0)+\tilde{\rho}(\vec{x},t)$.~The equation for $\vec{\xi}$ is 
\begin{align}
\frac{\partial^{2} \vec{\xi}}{\partial t^2}  + 2\eta \frac{\partial \vec{\xi}}{\partial t} + \mathcal{L}\vec{\xi} =  \vec{\mathcal{\vec{F}}},~\label{eq:xi_damping}
\end{align}
where $\eta$ is the damping rate, $\vec{\mathcal{F}}$ is a $\vx$-independent source term, and $\mathcal{L}$ is a second-order differential operator given by 
\begin{equation}
\label{eq:pulsation_operator} 
     \mathcal{L}\, \vec{\xi}=\frac{1}{\rho_0}\left(\vec{\nabla}\tilde{P}-\rho_0\tilde{\vec{g}}-\tilde{\rho}\vec{g}_0\right),
\end{equation}
where $P$ is the pressure and $\vec{g}=-\vec{\nabla}\Phi$ is the gravitational acceleration.~Each of the perturbed quantities can be related to $\vec{\xi}$ via the perturbed equations of stellar structure, which are the perturbed Poisson equation and the perturbed continuity and Euler equations \cite{aerts2010asteroseismology}, but we will not need these specific forms.~Equation~\eqref{eq:xi_damping} describes a damped, driven (anharmonic) oscillator.~Physically, the $\vF$ term encapsulates the  effects of turbulent convection, which acts to stochastically-excite stellar oscillations \cite{1977ApJ...212..243G,1980tsp..book.....C,2001A&A...370..136S,2001A&A...373..916L,2005MNRAS.360..859C}; and the term proportional to $\partial/\partial t $ represents the damping effects of turbulent viscosity \cite{1977ApJ...212..243G,2005MNRAS.360..859C}.~The frequency-dependent constant $\eta$ parametrizes the strength of the damping \cite{2005MNRAS.360..859C,Chaplin:2008af,2019MNRAS.487..595H}.

In the absence of damping ($\eta=0$) and stochastic excitation ($\vec{\mathcal{F}}=0$), equation~\eqref{eq:xi_damping} (in Fourier-space) is a Sturm-Liouville eigenvalue problem \cite{1980tsp..book.....C,1989nos..book.....U,aerts2010asteroseismology}
\begin{equation}
\label{eq:L_operator_def}
\mathcal{L}\vec{\xi}_{q}=\omega_q^2 \vec{\xi}_{q},
\end{equation}
for a series of eigenmodes $\vx_q$ with eigenfrequencies $\omega_q^2$.~It is common to expand the general solution as
\begin{equation}
    \vec{\xi}(\vec{r},t)=\sum_{n=0}^\infty\sum_{l=0}^\infty\sum_{m=-l}^l\vec{\xi}_{nlm}(\vec{r})e^{-i\omega_{nlm} t}
\end{equation}
with 
\begin{align}
    % \vec{\xi}_{q}~\equiv \vec{\xi}_{nlm}&=\sqrt{4\pi}\left[\xi^r_{nl}(r)Y_{lm}(\theta,\phi)\hat{\vec{r}}\nonumber\right.\\
    % &\left.+\xi^h_{nl}(r)\left(\frac{\partial Y_{lm}(\theta,\phi)}{\partial\theta}\hat{\vec{\theta}}+\frac{1}{\sin\theta}\frac{\partial Y_{lm}(\theta,\phi)}{\partial\phi}\hat{\vec{\phi}}\right)\right]\nonumb\\
    %%%%
    %&
    \vec{\xi}_{q} =\sqrt{4\pi}\left[\xi^r_{nl}(r)\vec{Y}_{lm}(\theta,\phi) + \xi^h_{nl}(r)\vec{\Psi}_{lm}(\theta,\phi) \right],
\label{eq:xi_in_vector_spherical_harmonics}
\end{align}
where $\vec{Y}_{lm}=Y_{lm}\hat{\vec{r}}$ and $\vec{\Psi}_{lm}=r\vec{\nabla}Y_{lm}$ are the vector spherical harmonics.~The eigenfunctions $\vx_q$ form a complete orthogonal basis with weight function $\rho_0$. We define the normalization as
\begin{equation}
\label{eq:eigenfunction_normalization}
\int\dd^3\vec{x}\,\rho(\vec{x}')\vec{\xi}_{q}^*(\vec{x})\cdot\vec{\xi}_{q'}(\vec{x})=I_q\delta_{qq'},
\end{equation}
where $I_q$ is the the mode's \textit{inertia}, which can be calculated given a solution for any particular eigenfunction.~Any given star will oscillate in a linear superposition of  eigenmodes $\vx_{nlm}$ with frequencies $\omega_{nlm}$.

To derive the effect of the oscillating Newtonian potential due to ULDM in equation~\eqref{eq:oscillating_potentials}, we can  write $\tilde{g}\rightarrow\tilde{g}-\nabla\Phi(\vec{x},t)$ to find a new source term in equation~\eqref{eq:xi_damping} $\vF_{\rm ULDM}= -\nabla\Phi(\vec{x},t)$.~The solution of~\eqref{eq:xi_damping} can be written as a sum over the eigenmodes of the form
\begin{equation}
\label{eq:velocity_field_expansion}
    \vec{\xi}(\vec{x},t)=\sum_{q}A_q(t)\vec{\xi}_q(\vec{x})e^{-i\omega_qt},
\end{equation}
where $A_{q}(t)$ are time-varying coefficients, which we take to be slowly-varying by imposing $\ddot{A} \ll \omega_{q}^2 A$ and $\eta \ll \omega_q$ (these approximations are known to be good for the objects we are interested in \cite{2005MNRAS.360..859C}).~Our goal is to determine $A_q(t)$.~Substituting equation~\eqref{eq:velocity_field_expansion} into~\eqref{eq:xi_damping}, multiplying both sides by the equilibrium stellar density profile $\rho_0(\vec{r})$ and conjugate eigenvector $\vec{\xi}^{*}_q(\vec{r})$ respectively and integrating throughout the star we find
\begin{equation}
\label{eq:amplitude_equations}
    \frac{\partial A_q}{\partial t}+\eta A_q = \frac{i}{2\omega_q I_q}e^{i\omega_q t}\mathcal{Q}_q(t),
\end{equation}
where we used the orthogonality condition of the mode functions~\eqref{eq:eigenfunction_normalization}, the homogeneous equation~\eqref{eq:L_operator_def}, and employed the slowly-varying approximations for $A_q(t)$ above.~The quantities
\begin{equation}
\label{eq:axion_forcing_term}
 \mathcal{Q}_q(t)=   
\int\dd^3\vec{x}\,\vec{\xi}^*_q(\vec{x})\cdot\vec{\mathcal{F}}_{\rm ULDM}(\vec{x},t)
\end{equation}
encode the effects of the ULDM forcing of the modes.~Note that we have neglected the stochastic sourcing of the modes due to turbulent convection.~We will return to this later.

We can evaluate $\Qq$ by writing $\rho_0\vec{\xi}^*_q\vec{\nabla}\Phi=\vec{\nabla}\cdot(\rho_0\Phi\vec{\vec{\xi}^*_q})-\Phi\vec{\nabla}\cdot(\rho_0\vec{\xi}^*_q) $.~The first terms is a total derivative so vanishes by virtue of the divergence theorem and we are left with
\begin{equation}
\label{eq:axion_forcing_term_intermediate}
     \mathcal{Q}_q(t)=   
    \int\dd^3\vec{x}\, \Phi(\vec{x})\vec{\nabla}\cdot\left[\rho_0(\vec{x})\vec{\xi}^*_q(\vec{x}))\right].
\end{equation}
Using equation~\eqref{eq:xi_in_vector_spherical_harmonics} we have\footnote{Two useful identities are \[\nabla\cdot(f\vec{Y}_{lm})=\left(\frac{\dd f}{\dd r}+\frac{2}{r}f\right)Y_{lm}\quad\textrm{and}\quad \nabla\cdot(f\vec{\Psi}_{lm}) =-\frac{l(l+1)}{r}fY_{lm}.~\]}
\begin{equation}
\label{eq:axion_forcing_general}
    \mathcal{Q}_q(t)={\sqrt{4\pi}}\int\dd^3\vec{x}\, \Phi(\vec{x},t)\mathfrak{f}_{nl}(r)Y^*_{lm}
\end{equation}
with
\begin{equation}
    \mathfrak{f}_{nl}(r)=\frac{\dd (\rho_0\xi_{nl}^r)}{\dd r}+\frac{2}{r}\rho_0\xi_{nl}^r-\frac{l(l+1)}{r}\rho_0\xi^h_{nl}.
\end{equation}
Equation~\eqref{eq:axion_forcing_general} is as far as we can go without specifying the potential so we now substitute equation~\eqref{eq:oscillating_potentials}.~Before doing this, we note that the velocity of the galaxy, and hence most stars, through the DM halo is non-relativistic.~Therefore, $v^\star/c\ll1$, and we can Taylor-expand the Newtonian potential to find 
\begin{align}
\label{eq:perturb_Phi}
   \Phi(\vec{x},t) &\simeq \frac{\pi G\gamma\omega\rho\dm}{m^2}  \vec{v}^\star\!\cdot\!\vec{x}\sin(\gamma\omega t)+\mathcal{O}((\vec{v}^\star\!\cdot\!\vec{x})^2),
    %\Phi(\vec{x},t) &\simeq -\frac{\pi G\rho\dm}{m^2}\left[- \gamma \omega \vec{v}^\star\cdot\vec{x}\sin(\gamma\omega t)+\mathcal{O}((\vec{v}^\star\cdot\vec{x})^2)\right],
\end{align}
where we dropped the space-independent zeroth-order term since it does not contribute to the Newtonian force.~For convenience, we take the direction of the stars's motion through the DM halo to be the $z$-axis, which implies that $\vec{v^\star}\cdot{\vec{x}}=vr\cos\theta$.~Equation~\eqref{eq:axion_forcing_general} is then 
\begin{align}
\mathcal{Q}_q(t)&=\frac{4\pi  G \rho\dm v^\star\gamma\omega}{\sqrt{3}m^2}\sin(\gamma\omega t)\int\dd r\, r^3 \,\mathfrak{f}_{nl}(r)\int\dd\Omega\, Y_{10} Y^*_{lm}\nonumber\\
    & = \bar{\mathcal{Q}}_q \sin(\gamma\omega t)\delta_{l1}\delta_{m 0}.
\end{align}
with
\begin{equation}
    \label{eq:W_definition}
\bar{\mathcal{Q}}_q\equiv\frac{8\pi^2 \mathcal{W}_{nl} G \rho\dm\gamma v^\star}{\sqrt{3}m};\quad\mathcal{W}_{nl}\equiv \int\dd r\, r^3 \,\mathfrak{f}_{nl}(r).
\end{equation}
Thus, we see that, at leading order in $v$, only the $l=1$ modes are excited by the ULDM-induced oscillating gravitational potential.~

Three comments are in order.~First, note that the $m=\pm1$ terms are not excited as a consequence of our choice of coordinate system, namely we chose the velocity to be aligned with the $z$-axis so that the spherical symmetry is broken to azimuthal symmetry.~This choice may not be optimal for stars where the rotational splitting of the modes are observable since it may then be preferential to take the $z$-axis to be aligned with the axis of rotation.~If a coordinate system is chosen such that $z$ and $\vec{v}$ are misaligned then all three $m$-modes will be excited.~Second, models of all order (including $l=0$) are excited by higher order terms in equation~\eqref{eq:perturb_Phi}, but they are suppressed by powers of $v^\star$.~In particular, at second-order the modes with $l=0$ and $l=2$ are excited.~Third, the evaluation of the solution of~\eqref{eq:amplitude_equations} for the amplitude of each mode requires the computation of two quantities which depend on the structure of the star.~These are $\mathcal{W}_{nl}$ and $I_p$, defined in equation~\eqref{eq:W_definition} and~\eqref{eq:eigenfunction_normalization} respectively.~ One must compute the equilibrium configuration for the star and use this to solve the adiabatic equation for the radial eigenfunctions.

We now proceed to solve equation~\eqref{eq:amplitude_equations}, which admits the solution 
\begin{equation}
\label{eq:amplitude_solution}
    A_q(t)=i \frac{\bar{\mathcal{Q}}}{2\omega_q I_q}e^{i\omega_q t}\left(\frac{ \gamma \omega \cos(\gamma \omega t)-(\eta+i\omega_q)\sin(\gamma\omega t)}{\gamma^2 \omega^2 + \eta^2-\omega_q^2+2i\omega_q\eta}\right)
\end{equation}
%In deriving the solution~\eqref{eq:amplitude_solution} we neglected the part of the solution which is $\propto e^{-\eta t}$, as it decays fast with time.~
Realistic observations are performed within a finite time-frame $T$.~Therefore, it is the time-averaged amplitude which is relevant, $\langle|A_q|^2\rangle$, and is defined as
\begin{equation}
\langle|A_q|^2\rangle~\equiv \frac{1}{T} \int_0^{T} dt |A_{q}(t)|^2, 
\end{equation}
with $T$ the observation time for the given mode $q$.~It is straightforward to show that 
\begin{equation}
    \label{eq:A_average}
    \langle|A_q|^2\rangle = \Big( \frac{\bar{\mathcal{Q}}_q^2}{4\omega_q^4I_q^2} \Big)\frac{\gamma^2 x^2 + 1}{(\gamma^2x^2 - 1)^2 + 4\eta^2/\omega_{q}^2},
\end{equation}
where $x=\omega/\omega_q$ and remind the reader that $\omega=2m$.~In deriving the above relation we used the approximation that $\eta \ll \omega_q, \omega$, which is valid in the typical objects of interest \cite{2005MNRAS.360..859C,2006ESASP.624E..28H,2019MNRAS.487..595H} \footnote{For example, the Sun's $n=0$, $l=1$ mode has a frequency of $\mathcal{O}(300\mu\textrm{Hz})$ and damping $\eta\sim 10^{-6}\mu$Hz.}, we we also set the time-averaged trigonometric functions equal to their approximate value $1/2$.~As we will see below, $ \langle|A_q|^2\rangle$ enters directly in the expression for the surface velocity amplitude.~It is clear that~\eqref{eq:A_average} assumes its maximum value when an exact resonance occurs, i.e in the limit $x \to 1$.~For practical computations we may set $\gamma \simeq 1$ since the velocity of the star through the DM halo is non-relativistic.~Equation~\eqref{eq:A_average} is our main result:~the ULDM-induced oscillating Newtonian potential resonantly-excites  stellar oscillation p-modes with degree $l=1$ (predominantly).

Stellar oscillations typically manifest as a surface velocity field or as variations in the brightness of the star.~Here, we will employ the former observable since in this work we will apply our formalism to the Sun, for which the induced surface velocity field associated with a wide range of frequencies has been measured with high precision.~The Sun is certainly the star with the most precise and accurate seismic observations, offering an ideal laboratory for our preliminary study.~The variation of the surface velocity field is measured through the induced Doppler shift of certain element emission lines along the line of sight.~The observed signal is a superposition of different stochastically-excited acoustic solar modes.~The space- and time-dependent observed velocity field along the line-of-sight, $V_{lmn}(\theta, \phi; t)$, undergoes a decomposition into spherical harmonics, from which individual modes of given $(l,m)$ are appropriately projected out.~The evolution of the velocity field is subsequently exploited to identify the overtones $n$ (see e.g.~\cite{aerts2010asteroseismology}, \cite{Christensen-Dalsgaard:2002ney}).

The observed solar acoustic oscillations are excited by turbulent motion of convective eddies in the outer parts of the Sun.~The strength of the damping is measured by the damping coefficient $\eta$ appearing in equation~\eqref{eq:xi_damping}, which is a frequency-dependent quantity because the damping depends on the characteristics of individual modes.~Its computation requires the solution of the non-adiabatic oscillation equations to account for the energy sinks and sources in the solar interior, and it receives contributions from different physical processes such as the scattering and viscous damping of interior acoustic waves.~The value and scaling of $\eta$ with solar frequencies has been modelled previously  e.g.~\cite{2019MNRAS.487..595H}.

We proceed with determining the effect of ULDM on the stellar surface velocity field.~Let us introduce the time-averaged kinetic energy of a given mode as
\begin{equation} \label{eq:kinetic_energy}
E^{(q)}_{\rm{kin}} = \frac{1}{T} \int_{0}^{T} dt \int_{\Omega} d\Omega \int_{0}^{R_{\odot}}  dr r^2 \, \rho_0(r) |\vec{v}_q(\vec{x},t)|^2, 
\end{equation}
where $\vec{v}_q(\vec{x},t)=\dot{\vx}_q$ is the mode's velocity, and $T$ denotes the observation time of the mode.~In order to evaluate this relation further, we note that the velocity field is related to the displacement vector via $\vec{v}_q(\vec{x},t)\simeq \omega_{q}^2 |\vec{\xi}|^2$
% \begin{equation}
%     |\vec{v}(\vec{x},t)|^2= \left| \frac{\partial{\vec{\xi}}}{\partial t} \right|^2 \simeq \omega_{q}^2 |\vec{\xi}|^2, 
% \end{equation}
where we expanded $\vx$ as $ \vec{\xi} = \sum_qA_q(t) \vec{\xi}_{q}(\vec{x}) e^{-i\omega_qt}$ and used the approximation of a slowly-varying amplitude, i.e.,~$\partial \ln A/\partial t \ll \omega_q$.~Substituting into~\eqref{eq:kinetic_energy} and time-averaging we find
\begin{align}
\label{eq:Ekin_def1}
E^{(q)}_{\rm{kin}} & = \frac{1}{2} \langle |A_q(t)|^2 \rangle \omega_{q}^2 I_q.
\end{align}
%where $I_q$ is given by \js{should we remove this sentence and the equation below? We already define $I_q$ above and we could save space for PRL.~If we do keep it maybe just the lase equality since this is new.} \is{You can remove the equation for $I_q$.~I don't understand which sentence you mean.}
% \begin{align}
% I_q = \!\int \dd r r^2 \rho_0 |\vec{\xi}_q|^2
%  =  4 \pi \!\!\int \dd r r^2 \rho_0 \!\left( \xi_{r}^{2} + l(l+1)\xi_{h}^{2} \right).
% \end{align}
Now, and in analogy with the standard kinetic energy of a particle, the kinetic energy of a pulsation mode $q$ can be expressed in terms of the root-mean-squared (RMS) surface velocity $V_{\rm{rms}}$ associated with this mode (we neglect the index $q$ for simplicity) as 
\begin{equation}
\label{eq:Ekin_def2}
E_{\rm{kin}}^{(q)} = \frac{1}{2} M_{q} V^{2}_{\rm{rms}}, 
\end{equation}
with the mode's mass defined as $M_{q}~\equiv {I_{q}(r)}/{|\vec{\xi}|^2_{r = R_{\odot}}}$.~It is understood that the velocity RMS above is the one associated with each mode $q$.~% \begin{equation}
% M_{q}~\equiv \frac{I_{q}(r)}{|\vec{\xi}|^2_{r = R_{\odot}}}.
% \end{equation}
Comparing equations~\eqref{eq:Ekin_def1} and~\eqref{eq:Ekin_def2} one can derive the following expression for the RMS velocity of the mode,
\begin{equation} \label{eq:Vrms}
V^{2}_{\rm{rms}} = |\vec{\xi}_q|^2_{r = R_{\odot}} \omega^{2}_{q}  \langle |A_q(t)|^2 \rangle.
\end{equation}
%Clearly, $V_{\rm{rms}}$ depends on the frequency of the  mode.

Before calculating the amplitude of $V_{\rm rms}$ predicted by ULDM, we address the stochastic sourcing of the modes by turbulent convection, which we have thus far ignored.~The amplitude equation~\eqref{eq:xi_damping} has two forcing terms, one due to turbulence and another due to ULDM.~The solution for the former is very challenging due to the stochastic nature of turbulent excitations, although significant progress can be made under certain approximations (see e.g.~\cite{Kolmogorov1890}).~The linearity of equation~\eqref{eq:xi_damping} allows us to treat the total solution for the amplitude as the linear combination of the individual solutions derived with only one of the source terms present at a time, that is, $A(t)^{\rm (tot.)} = A(t)^{\rm (conv.)} + A(t)^{\rm (ULDM)}$.~The total average squared-amplitude is then 
\begin{align} \label{eq:A-average-total}
\langle |A^{\rm (tot.)}|^2 \rangle & \; = \; \langle |A^{\rm (conv.)}|^2 \rangle + \langle |A^{\rm (ULDM)}|^2 \rangle  \nonumber \\ 
& + 2 \langle \mathrm{Re}{(A^{\rm (conv.)} A^{\rm (ULDM)})} \rangle.
\end{align}
The ULDM resonance is observable if $\langle|A^{\rm (ULDM)}|^2 \rangle>\langle |A^{\rm (conv.)}|^2 \rangle$ provided that the cross-term is negligible.~To estimate the cross-term we need an estimate for $A^{\rm (conv.)}$, which is given by \cite{Samadi:2001nc}
\begin{equation}
    A^{\rm (conv.)}(t) \simeq  \frac{ie^{i\omega_q t}}{2\omega_qI_q}\int\dd^3\vec{x}\,\vec{\xi}^*(\vec{x})\cdot\vec{\mathcal{S}(\vec{x})},
\end{equation}
where $\mathcal{S}(\vec{x})$ is a form factor containing the stochastic effects of turbulent convection.~We have assumed that $\mathcal{S}(\vec{x})$ is approximately time-independent by virtue of the difference between the timescales of convection and solar oscillations ($t_{\rm osc.} \ll t_{\rm conv.}\sim\eta^{-1}$).~We therefore have $\mathrm{Re}{(A^{\rm (conv.)} A^{\rm (ULDM)})} \rangle\propto \langle\sin(\gamma\omega t)\rangle,\,\langle\cos(\gamma\omega t)\rangle=0$.~From~\eqref{eq:Vrms} it then follows for the total surface velocity amplitude is the sum of the individual contributions from convection and ULDM
\begin{equation}
V^{2}_{\rm{rms}} \simeq {V^{2}_{\rm{rms}}}^{\rm (conv.)} + {V^{2}_{\rm{rms}}}^{\rm (ULDM)}.
\end{equation}
%Therefore, the total surface velocity squared amplitude will be given by the linear sum of that contributed by the convection and ULDM excitations respectively.~

To estimate the size of $V_{\rm{rms}}^{\rm (ULDM)}$, we can take the limit $\gamma=1$, $x=1$, and $\omega_q=2m$ i.e., exact resonance in equations~\eqref{eq:A_average} and~\eqref{eq:Vrms}, in which case we find
\begin{widetext}
\begin{equation}
\label{eq:VRMS_ULDM_estimate}
    V_{\rm{rms}}^{\rm (ULDM)} \simeq 3.3\times10^{-8}\left(\frac{\rho\dm}{0.42\textrm{GeV/cm}^3}\right)\left(\frac{v^\star}{220\textrm{km/s}}\right)\left(\frac{150\mu\textrm{Hz}}{m}\right)\left(\frac{2\times10^{-6}\mu\textrm{Hz}}{\eta}\right)\left(\frac{\mathcal{W}_{nl}|\vec{\xi}_q|_{r = R_{\odot}} }{I_q}\right)\textrm{cm/s},
\end{equation}
\end{widetext}
where the fiducial values  correspond to the local DM density in the solar system, the circular velocity of the Milky Way at the location of the solar system, half the frequency of the solar mode with $n=1$, $l=1$, and the damping rate of this mode.\footnote{We are grateful to G.~Houdek for poviding us with this value.} 

A typical solar surface velocity amplitude is $\mathcal{O}(1\textrm{cm/s})$ \cite{GoudekPhD, Libbrecht1998}, so whether or not the ULDM-induced mode excitations are detectable depends crucially on the dimensionless form factor $\mathcal{W}_{nl}|\vec{\xi}_q|_{r = R_{\odot}}/I_q$.~To calculate this, we calibrated an equilibrium solar model using the stellar-evolution code MESA \cite{Paxton2011,Paxton2013,Paxton2015,Paxton2018,Paxton2019}, which we subsequently used to compute the oscillation eigenspectrum using the stellar oscillation code GYRE \cite{2013MNRAS.435.3406T}.~This model --- created and discussed in  \cite{Saltas:2022ybg} --- reproduces the surface metallicity, luminosity, and radius of the present Sun within $1\sigma$ of the respective observational errors.~The model has initial metallicity $Z_{\rm{in}} = 0.0186$, initial Helium abundance $Y_{\rm{in}} = 0.2690$, and mixing-length parameter $\alpha_{\rm{MLT}} = 2.0031$; and assumes a GS98 metal mixture \cite{GS98}, an OPAL equation of state \cite{Rogers2002}, and the opacity tabulation of OP \cite{Seaton2005}.

The mode with $\{n,\,l\}=\{1,\,1\}$ has frequency $\omega_{1,1}=285\mu$Hz, and we found $\mathcal{W}_{nl}|\vec{\xi}_q|_{r = R_{\odot}}/I_q=0.6$ implying that $V_{\rm{rms}}^{\rm (ULDM)} \simeq2\times10^{-8}$cm/s.~We found that the form factor for higher $n$ modes was not significantly different, meanwhile the frequency and damping rates are significantly larger (e.g., $\omega=3650\mu$Hz and $\eta\sim10\mu$Hz \cite{2006ESASP.624E..28H} for $n=25$).~We therefore conclude that ULDM-induced solar oscillations are likely undetectable.~

In light of this conclusion, it is prudent to examine the prospects for observing ULDM-induced oscillations in other objects.~Equation~\eqref{eq:VRMS_ULDM_estimate} implies that solar-like oscillators in environments where the DM density is higher i.e., closer to the center of the galaxy are more strongly excited by ULDM.~The DM density near the center of the galaxy is $\rho\dm\sim840$ GeV/cm$^3$ \cite{2020Galax...8...37S}, which, when applied to equation~\eqref{eq:VRMS_ULDM_estimate} still produces a negligible RMS surface velocity.~Given that the velocity $v^\star$ is likely smaller closer to the galactic center on account of the smaller circular velocity, and the fact that solar-like oscillators are not presently observed close to the galactic center, the study of other objects will likely prove more fruitful.~Clearly, the resonance effect would be amplified for a smaller damping rate.~Taking into account the scaling of $\eta$ with effective temperature ($T_{\rm eff}$) and compactness ($g$), $\eta \propto T_{\rm eff}^{10.8} g^{-0.3}$ \cite{Belkacem2012}, it seems that promising targets would be stars with near-solar compactness, but with much lower temperatures, such as brown dwarfs or red giants. Seismic observations for brown dwarfs might be challenging due to their faintness, but red giants (and their oscillations) are observationally abundant throughout the galaxy \cite{2010ApJ...723.1607H,2018ApJS..236...42Y,2019MNRAS.485.5616H,2022A&A...667A..31B,2022MNRAS.512.1677S,2022MNRAS.512.1677S,2022AJ....164..135H,2020ApJ...889L..34S}.~Finally, it would be interesting to determine if the resonant excitation of neutron star oscillations by ULDM could lead to observable gravitational wave signatures.~Such a study would require our formalism to be extended to relativistic stars \cite{1967ApJ...149..591T,1991RSPSA.432..247C,B_F_Schutz_2008,Kokkotas:1999bd}.~These objects are relativistic and are therefore sensitive to the oscillating ULDM pressure so are expected exhibit effects even in the case $v^\star\rightarrow0$.

To summarize, we have identified a novel effect of ULDM models whereby stellar oscillation modes can be resonantly-excited by the oscillating Newtonian potential sourced by the DM's coherent harmomic motion.~We developed a mathematical framework for predicting the amplitude of the root-mean-square surface velocity amplitude due to this excitation that is appropriate for linear, adiabatic stellar oscillations.~Applying this to a model for the Sun, we found that ULDM-induced solar oscillations are likely undetectable.~We discussed several avenues for followup work.

\textit{Acknowledgements.} We are grateful for discussions with Djuna Croon, J\o rgen Christensen-Dalsgaard, G\"{u}nther Houdek, Daniel Huber, and Joel Ong.~This work was initiated at the workshop \textit{Stellar Tests of Gravity}, held at Carnegie Mellon University in March 2022. IDS received support by the Czech Grant Agency (GA\^CR) under the grant number 21-16583M.

\textit{Software.} MESA version 15140, Python version 3.9.7, GYRE version 7.0.~A reproduction package containing our inlists and models, and our Python notebook for calculating the form factor in equation~\eqref{eq:VRMS_ULDM_estimate} is available at the following URL \cite{sakstein_jeremy_2023_7894030}: \href{https://zenodo.org/record/7894030}{https://zenodo.org/record/7894030}.

\bibliography{sample}

\begin{thebibliography}{59}%
\makeatletter
\providecommand \@ifxundefined [1]{%
 \@ifx{#1\undefined}
}%
\providecommand \@ifnum [1]{%
 \ifnum #1\expandafter \@firstoftwo
 \else \expandafter \@secondoftwo
 \fi
}%
\providecommand \@ifx [1]{%
 \ifx #1\expandafter \@firstoftwo
 \else \expandafter \@secondoftwo
 \fi
}%
\providecommand \natexlab [1]{#1}%
\providecommand \enquote  [1]{``#1''}%
\providecommand \bibnamefont  [1]{#1}%
\providecommand \bibfnamefont [1]{#1}%
\providecommand \citenamefont [1]{#1}%
\providecommand \href@noop [0]{\@secondoftwo}%
\providecommand \href [0]{\begingroup \@sanitize@url \@href}%
\providecommand \@href[1]{\@@startlink{#1}\@@href}%
\providecommand \@@href[1]{\endgroup#1\@@endlink}%
\providecommand \@sanitize@url [0]{\catcode `\\12\catcode `\$12\catcode
  `\&12\catcode `\#12\catcode `\^12\catcode `\_12\catcode `\%12\relax}%
\providecommand \@@startlink[1]{}%
\providecommand \@@endlink[0]{}%
\providecommand \url  [0]{\begingroup\@sanitize@url \@url }%
\providecommand \@url [1]{\endgroup\@href {#1}{\urlprefix }}%
\providecommand \urlprefix  [0]{URL }%
\providecommand \Eprint [0]{\href }%
\providecommand \doibase [0]{https://doi.org/}%
\providecommand \selectlanguage [0]{\@gobble}%
\providecommand \bibinfo  [0]{\@secondoftwo}%
\providecommand \bibfield  [0]{\@secondoftwo}%
\providecommand \translation [1]{[#1]}%
\providecommand \BibitemOpen [0]{}%
\providecommand \bibitemStop [0]{}%
\providecommand \bibitemNoStop [0]{.\EOS\space}%
\providecommand \EOS [0]{\spacefactor3000\relax}%
\providecommand \BibitemShut  [1]{\csname bibitem#1\endcsname}%
\let\auto@bib@innerbib\@empty
%</preamble>
\bibitem [{\citenamefont {Marsh}(2016)}]{Marsh:2015xka}%
  \BibitemOpen
  \bibfield  {author} {\bibinfo {author} {\bibfnamefont {D.~J.~E.}\
  \bibnamefont {Marsh}},\ }\bibfield  {title} {\bibinfo {title} {{Axion
  Cosmology}},\ }\href {https://doi.org/10.1016/j.physrep.2016.06.005}
  {\bibfield  {journal} {\bibinfo  {journal} {Phys. Rept.}\ }\textbf {\bibinfo
  {volume} {643}},\ \bibinfo {pages} {1} (\bibinfo {year} {2016})},\ \Eprint
  {https://arxiv.org/abs/1510.07633} {arXiv:1510.07633 [astro-ph.CO]}
  \BibitemShut {NoStop}%
\bibitem [{\citenamefont {Hui}\ \emph {et~al.}(2017)\citenamefont {Hui},
  \citenamefont {Ostriker}, \citenamefont {Tremaine},\ and\ \citenamefont
  {Witten}}]{Hui:2016ltb}%
  \BibitemOpen
  \bibfield  {author} {\bibinfo {author} {\bibfnamefont {L.}~\bibnamefont
  {Hui}}, \bibinfo {author} {\bibfnamefont {J.~P.}\ \bibnamefont {Ostriker}},
  \bibinfo {author} {\bibfnamefont {S.}~\bibnamefont {Tremaine}},\ and\
  \bibinfo {author} {\bibfnamefont {E.}~\bibnamefont {Witten}},\ }\bibfield
  {title} {\bibinfo {title} {{Ultralight scalars as cosmological dark
  matter}},\ }\href {https://doi.org/10.1103/PhysRevD.95.043541} {\bibfield
  {journal} {\bibinfo  {journal} {Phys. Rev. D}\ }\textbf {\bibinfo {volume}
  {95}},\ \bibinfo {pages} {043541} (\bibinfo {year} {2017})},\ \Eprint
  {https://arxiv.org/abs/1610.08297} {arXiv:1610.08297 [astro-ph.CO]}
  \BibitemShut {NoStop}%
\bibitem [{\citenamefont {Ferreira}(2021)}]{Ferreira:2020fam}%
  \BibitemOpen
  \bibfield  {author} {\bibinfo {author} {\bibfnamefont {E.~G.~M.}\
  \bibnamefont {Ferreira}},\ }\bibfield  {title} {\bibinfo {title}
  {{Ultra-light dark matter}},\ }\href
  {https://doi.org/10.1007/s00159-021-00135-6} {\bibfield  {journal} {\bibinfo
  {journal} {Astron. Astrophys. Rev.}\ }\textbf {\bibinfo {volume} {29}},\
  \bibinfo {pages} {7} (\bibinfo {year} {2021})},\ \Eprint
  {https://arxiv.org/abs/2005.03254} {arXiv:2005.03254 [astro-ph.CO]}
  \BibitemShut {NoStop}%
\bibitem [{\citenamefont {Hui}(2021)}]{Hui:2021tkt}%
  \BibitemOpen
  \bibfield  {author} {\bibinfo {author} {\bibfnamefont {L.}~\bibnamefont
  {Hui}},\ }\bibfield  {title} {\bibinfo {title} {{Wave Dark Matter}},\ }\href
  {https://doi.org/10.1146/annurev-astro-120920-010024} {\bibfield  {journal}
  {\bibinfo  {journal} {Ann. Rev. Astron. Astrophys.}\ }\textbf {\bibinfo
  {volume} {59}},\ \bibinfo {pages} {247} (\bibinfo {year} {2021})},\ \Eprint
  {https://arxiv.org/abs/2101.11735} {arXiv:2101.11735 [astro-ph.CO]}
  \BibitemShut {NoStop}%
\bibitem [{\citenamefont {Chadha-Day}\ \emph {et~al.}(2022)\citenamefont
  {Chadha-Day}, \citenamefont {Ellis},\ and\ \citenamefont
  {Marsh}}]{Chadha-Day:2021szb}%
  \BibitemOpen
  \bibfield  {author} {\bibinfo {author} {\bibfnamefont {F.}~\bibnamefont
  {Chadha-Day}}, \bibinfo {author} {\bibfnamefont {J.}~\bibnamefont {Ellis}},\
  and\ \bibinfo {author} {\bibfnamefont {D.~J.~E.}\ \bibnamefont {Marsh}},\
  }\bibfield  {title} {\bibinfo {title} {{Axion dark matter: What is it and why
  now?}},\ }\href {https://doi.org/10.1126/sciadv.abj3618} {\bibfield
  {journal} {\bibinfo  {journal} {Sci. Adv.}\ }\textbf {\bibinfo {volume}
  {8}},\ \bibinfo {pages} {abj3618} (\bibinfo {year} {2022})},\ \Eprint
  {https://arxiv.org/abs/2105.01406} {arXiv:2105.01406 [hep-ph]} \BibitemShut
  {NoStop}%
\bibitem [{\citenamefont {Hu}\ \emph {et~al.}(2000)\citenamefont {Hu},
  \citenamefont {Barkana},\ and\ \citenamefont {Gruzinov}}]{Hu:2000ke}%
  \BibitemOpen
  \bibfield  {author} {\bibinfo {author} {\bibfnamefont {W.}~\bibnamefont
  {Hu}}, \bibinfo {author} {\bibfnamefont {R.}~\bibnamefont {Barkana}},\ and\
  \bibinfo {author} {\bibfnamefont {A.}~\bibnamefont {Gruzinov}},\ }\bibfield
  {title} {\bibinfo {title} {{Cold and fuzzy dark matter}},\ }\href
  {https://doi.org/10.1103/PhysRevLett.85.1158} {\bibfield  {journal} {\bibinfo
   {journal} {Phys. Rev. Lett.}\ }\textbf {\bibinfo {volume} {85}},\ \bibinfo
  {pages} {1158} (\bibinfo {year} {2000})},\ \Eprint
  {https://arxiv.org/abs/astro-ph/0003365} {arXiv:astro-ph/0003365}
  \BibitemShut {NoStop}%
\bibitem [{\citenamefont {Peccei}\ and\ \citenamefont
  {Quinn}(1977)}]{Peccei:1977hh}%
  \BibitemOpen
  \bibfield  {author} {\bibinfo {author} {\bibfnamefont {R.~D.}\ \bibnamefont
  {Peccei}}\ and\ \bibinfo {author} {\bibfnamefont {H.~R.}\ \bibnamefont
  {Quinn}},\ }\bibfield  {title} {\bibinfo {title} {{CP Conservation in the
  Presence of Instantons}},\ }\href
  {https://doi.org/10.1103/PhysRevLett.38.1440} {\bibfield  {journal} {\bibinfo
   {journal} {Phys. Rev. Lett.}\ }\textbf {\bibinfo {volume} {38}},\ \bibinfo
  {pages} {1440} (\bibinfo {year} {1977})}\BibitemShut {NoStop}%
\bibitem [{\citenamefont {Weinberg}(1978)}]{Weinberg:1977ma}%
  \BibitemOpen
  \bibfield  {author} {\bibinfo {author} {\bibfnamefont {S.}~\bibnamefont
  {Weinberg}},\ }\bibfield  {title} {\bibinfo {title} {{A New Light Boson?}},\
  }\href {https://doi.org/10.1103/PhysRevLett.40.223} {\bibfield  {journal}
  {\bibinfo  {journal} {Phys. Rev. Lett.}\ }\textbf {\bibinfo {volume} {40}},\
  \bibinfo {pages} {223} (\bibinfo {year} {1978})}\BibitemShut {NoStop}%
\bibitem [{\citenamefont {Wilczek}(1978)}]{Wilczek:1977pj}%
  \BibitemOpen
  \bibfield  {author} {\bibinfo {author} {\bibfnamefont {F.}~\bibnamefont
  {Wilczek}},\ }\bibfield  {title} {\bibinfo {title} {{Problem of Strong $P$
  and $T$ Invariance in the Presence of Instantons}},\ }\href
  {https://doi.org/10.1103/PhysRevLett.40.279} {\bibfield  {journal} {\bibinfo
  {journal} {Phys. Rev. Lett.}\ }\textbf {\bibinfo {volume} {40}},\ \bibinfo
  {pages} {279} (\bibinfo {year} {1978})}\BibitemShut {NoStop}%
\bibitem [{\citenamefont {Zhitnitsky}(1980)}]{Zhitnitsky:1980tq}%
  \BibitemOpen
  \bibfield  {author} {\bibinfo {author} {\bibfnamefont {A.~R.}\ \bibnamefont
  {Zhitnitsky}},\ }\bibfield  {title} {\bibinfo {title} {{On Possible
  Suppression of the Axion Hadron Interactions. (In Russian)}},\ }\href@noop {}
  {\bibfield  {journal} {\bibinfo  {journal} {Sov. J. Nucl. Phys.}\ }\textbf
  {\bibinfo {volume} {31}},\ \bibinfo {pages} {260} (\bibinfo {year}
  {1980})}\BibitemShut {NoStop}%
\bibitem [{\citenamefont {Svrcek}\ and\ \citenamefont
  {Witten}(2006)}]{Svrcek:2006yi}%
  \BibitemOpen
  \bibfield  {author} {\bibinfo {author} {\bibfnamefont {P.}~\bibnamefont
  {Svrcek}}\ and\ \bibinfo {author} {\bibfnamefont {E.}~\bibnamefont
  {Witten}},\ }\bibfield  {title} {\bibinfo {title} {{Axions In String
  Theory}},\ }\href {https://doi.org/10.1088/1126-6708/2006/06/051} {\bibfield
  {journal} {\bibinfo  {journal} {JHEP}\ }\textbf {\bibinfo {volume} {06}},\
  \bibinfo {pages} {051}},\ \Eprint {https://arxiv.org/abs/hep-th/0605206}
  {arXiv:hep-th/0605206} \BibitemShut {NoStop}%
\bibitem [{\citenamefont {Arvanitaki}\ \emph {et~al.}(2010)\citenamefont
  {Arvanitaki}, \citenamefont {Dimopoulos}, \citenamefont {Dubovsky},
  \citenamefont {Kaloper},\ and\ \citenamefont
  {March-Russell}}]{Arvanitaki:2009fg}%
  \BibitemOpen
  \bibfield  {author} {\bibinfo {author} {\bibfnamefont {A.}~\bibnamefont
  {Arvanitaki}}, \bibinfo {author} {\bibfnamefont {S.}~\bibnamefont
  {Dimopoulos}}, \bibinfo {author} {\bibfnamefont {S.}~\bibnamefont
  {Dubovsky}}, \bibinfo {author} {\bibfnamefont {N.}~\bibnamefont {Kaloper}},\
  and\ \bibinfo {author} {\bibfnamefont {J.}~\bibnamefont {March-Russell}},\
  }\bibfield  {title} {\bibinfo {title} {{String Axiverse}},\ }\href
  {https://doi.org/10.1103/PhysRevD.81.123530} {\bibfield  {journal} {\bibinfo
  {journal} {Phys. Rev. D}\ }\textbf {\bibinfo {volume} {81}},\ \bibinfo
  {pages} {123530} (\bibinfo {year} {2010})},\ \Eprint
  {https://arxiv.org/abs/0905.4720} {arXiv:0905.4720 [hep-th]} \BibitemShut
  {NoStop}%
\bibitem [{\citenamefont {Berezhiani}\ and\ \citenamefont
  {Khoury}(2015)}]{Berezhiani:2015bqa}%
  \BibitemOpen
  \bibfield  {author} {\bibinfo {author} {\bibfnamefont {L.}~\bibnamefont
  {Berezhiani}}\ and\ \bibinfo {author} {\bibfnamefont {J.}~\bibnamefont
  {Khoury}},\ }\bibfield  {title} {\bibinfo {title} {{Theory of dark matter
  superfluidity}},\ }\href {https://doi.org/10.1103/PhysRevD.92.103510}
  {\bibfield  {journal} {\bibinfo  {journal} {Phys. Rev. D}\ }\textbf {\bibinfo
  {volume} {92}},\ \bibinfo {pages} {103510} (\bibinfo {year} {2015})},\
  \Eprint {https://arxiv.org/abs/1507.01019} {arXiv:1507.01019 [astro-ph.CO]}
  \BibitemShut {NoStop}%
\bibitem [{\citenamefont {Khoury}(2022)}]{Khoury:2021tvy}%
  \BibitemOpen
  \bibfield  {author} {\bibinfo {author} {\bibfnamefont {J.}~\bibnamefont
  {Khoury}},\ }\bibfield  {title} {\bibinfo {title} {{Dark Matter
  Superfluidity}},\ }\href {https://doi.org/10.21468/SciPostPhysLectNotes.42}
  {\bibfield  {journal} {\bibinfo  {journal} {SciPost Phys. Lect. Notes}\
  }\textbf {\bibinfo {volume} {42}},\ \bibinfo {pages} {1} (\bibinfo {year}
  {2022})},\ \Eprint {https://arxiv.org/abs/2109.10928} {arXiv:2109.10928
  [astro-ph.CO]} \BibitemShut {NoStop}%
\bibitem [{\citenamefont {Khmelnitsky}\ and\ \citenamefont
  {Rubakov}(2014)}]{Khmelnitsky:2013lxt}%
  \BibitemOpen
  \bibfield  {author} {\bibinfo {author} {\bibfnamefont {A.}~\bibnamefont
  {Khmelnitsky}}\ and\ \bibinfo {author} {\bibfnamefont {V.}~\bibnamefont
  {Rubakov}},\ }\bibfield  {title} {\bibinfo {title} {{Pulsar timing signal
  from ultralight scalar dark matter}},\ }\href
  {https://doi.org/10.1088/1475-7516/2014/02/019} {\bibfield  {journal}
  {\bibinfo  {journal} {JCAP}\ }\textbf {\bibinfo {volume} {02}},\ \bibinfo
  {pages} {019}},\ \Eprint {https://arxiv.org/abs/1309.5888} {arXiv:1309.5888
  [astro-ph.CO]} \BibitemShut {NoStop}%
\bibitem [{\citenamefont {Aoki}\ and\ \citenamefont
  {Soda}(2016)}]{Aoki:2016kwl}%
  \BibitemOpen
  \bibfield  {author} {\bibinfo {author} {\bibfnamefont {A.}~\bibnamefont
  {Aoki}}\ and\ \bibinfo {author} {\bibfnamefont {J.}~\bibnamefont {Soda}},\
  }\bibfield  {title} {\bibinfo {title} {{Detecting ultralight axion dark
  matter wind with laser interferometers}},\ }\href
  {https://doi.org/10.1142/S0218271817500638} {\bibfield  {journal} {\bibinfo
  {journal} {Int. J. Mod. Phys. D}\ }\textbf {\bibinfo {volume} {26}},\
  \bibinfo {pages} {1750063} (\bibinfo {year} {2016})},\ \Eprint
  {https://arxiv.org/abs/1608.05933} {arXiv:1608.05933 [astro-ph.CO]}
  \BibitemShut {NoStop}%
\bibitem [{\citenamefont {Blas}\ \emph {et~al.}(2017)\citenamefont {Blas},
  \citenamefont {Nacir},\ and\ \citenamefont {Sibiryakov}}]{Blas:2016ddr}%
  \BibitemOpen
  \bibfield  {author} {\bibinfo {author} {\bibfnamefont {D.}~\bibnamefont
  {Blas}}, \bibinfo {author} {\bibfnamefont {D.~L.}\ \bibnamefont {Nacir}},\
  and\ \bibinfo {author} {\bibfnamefont {S.}~\bibnamefont {Sibiryakov}},\
  }\bibfield  {title} {\bibinfo {title} {{Ultralight Dark Matter Resonates with
  Binary Pulsars}},\ }\href {https://doi.org/10.1103/PhysRevLett.118.261102}
  {\bibfield  {journal} {\bibinfo  {journal} {Phys. Rev. Lett.}\ }\textbf
  {\bibinfo {volume} {118}},\ \bibinfo {pages} {261102} (\bibinfo {year}
  {2017})},\ \Eprint {https://arxiv.org/abs/1612.06789} {arXiv:1612.06789
  [hep-ph]} \BibitemShut {NoStop}%
\bibitem [{\citenamefont {Blas}\ \emph {et~al.}(2020)\citenamefont {Blas},
  \citenamefont {L\'opez~Nacir},\ and\ \citenamefont
  {Sibiryakov}}]{Blas:2019hxz}%
  \BibitemOpen
  \bibfield  {author} {\bibinfo {author} {\bibfnamefont {D.}~\bibnamefont
  {Blas}}, \bibinfo {author} {\bibfnamefont {D.}~\bibnamefont
  {L\'opez~Nacir}},\ and\ \bibinfo {author} {\bibfnamefont {S.}~\bibnamefont
  {Sibiryakov}},\ }\bibfield  {title} {\bibinfo {title} {{Secular effects of
  ultralight dark matter on binary pulsars}},\ }\href
  {https://doi.org/10.1103/PhysRevD.101.063016} {\bibfield  {journal} {\bibinfo
   {journal} {Phys. Rev. D}\ }\textbf {\bibinfo {volume} {101}},\ \bibinfo
  {pages} {063016} (\bibinfo {year} {2020})},\ \Eprint
  {https://arxiv.org/abs/1910.08544} {arXiv:1910.08544 [gr-qc]} \BibitemShut
  {NoStop}%
\bibitem [{\citenamefont {{Cox}}(1980)}]{1980tsp..book.....C}%
  \BibitemOpen
  \bibfield  {author} {\bibinfo {author} {\bibfnamefont {J.~P.}\ \bibnamefont
  {{Cox}}},\ }\href@noop {} {\emph {\bibinfo {title} {{Theory of Stellar
  Pulsation. (PSA-2), Volume 2}}}},\ Vol.~\bibinfo {volume} {2}\ (\bibinfo
  {year} {1980})\BibitemShut {NoStop}%
\bibitem [{\citenamefont {{Unno}}\ \emph {et~al.}(1989)\citenamefont {{Unno}},
  \citenamefont {{Osaki}}, \citenamefont {{Ando}}, \citenamefont {{Saio}},\
  and\ \citenamefont {{Shibahashi}}}]{1989nos..book.....U}%
  \BibitemOpen
  \bibfield  {author} {\bibinfo {author} {\bibfnamefont {W.}~\bibnamefont
  {{Unno}}}, \bibinfo {author} {\bibfnamefont {Y.}~\bibnamefont {{Osaki}}},
  \bibinfo {author} {\bibfnamefont {H.}~\bibnamefont {{Ando}}}, \bibinfo
  {author} {\bibfnamefont {H.}~\bibnamefont {{Saio}}},\ and\ \bibinfo {author}
  {\bibfnamefont {H.}~\bibnamefont {{Shibahashi}}},\ }\href@noop {} {\emph
  {\bibinfo {title} {{Nonradial oscillations of stars}}}}\ (\bibinfo {year}
  {1989})\BibitemShut {NoStop}%
\bibitem [{\citenamefont {{Samadi}}\ and\ \citenamefont
  {{Goupil}}(2001)}]{2001A&A...370..136S}%
  \BibitemOpen
  \bibfield  {author} {\bibinfo {author} {\bibfnamefont {R.}~\bibnamefont
  {{Samadi}}}\ and\ \bibinfo {author} {\bibfnamefont {M.~J.}\ \bibnamefont
  {{Goupil}}},\ }\bibfield  {title} {\bibinfo {title} {{Excitation of stellar
  p-modes by turbulent convection. I. Theoretical formulation}},\ }\href
  {https://doi.org/10.1051/0004-6361:20010212} {\bibfield  {journal} {\bibinfo
  {journal} {\aap}\ }\textbf {\bibinfo {volume} {370}},\ \bibinfo {pages} {136}
  (\bibinfo {year} {2001})},\ \Eprint {https://arxiv.org/abs/astro-ph/0101109}
  {arXiv:astro-ph/0101109 [astro-ph]} \BibitemShut {NoStop}%
\bibitem [{\citenamefont {{Lopes}}(2001)}]{2001A&A...373..916L}%
  \BibitemOpen
  \bibfield  {author} {\bibinfo {author} {\bibfnamefont {I.~P.}\ \bibnamefont
  {{Lopes}}},\ }\bibfield  {title} {\bibinfo {title} {{Nonradial adiabatic
  oscillations of stars. Mode classification of acoustic-gravity waves}},\
  }\href {https://doi.org/10.1051/0004-6361:20010130} {\bibfield  {journal}
  {\bibinfo  {journal} {\aap}\ }\textbf {\bibinfo {volume} {373}},\ \bibinfo
  {pages} {916} (\bibinfo {year} {2001})}\BibitemShut {NoStop}%
\bibitem [{\citenamefont {{Chaplin}}\ \emph {et~al.}(2005)\citenamefont
  {{Chaplin}}, \citenamefont {{Houdek}}, \citenamefont {{Elsworth}},
  \citenamefont {{Gough}}, \citenamefont {{Isaak}},\ and\ \citenamefont
  {{New}}}]{2005MNRAS.360..859C}%
  \BibitemOpen
  \bibfield  {author} {\bibinfo {author} {\bibfnamefont {W.~J.}\ \bibnamefont
  {{Chaplin}}}, \bibinfo {author} {\bibfnamefont {G.}~\bibnamefont {{Houdek}}},
  \bibinfo {author} {\bibfnamefont {Y.}~\bibnamefont {{Elsworth}}}, \bibinfo
  {author} {\bibfnamefont {D.~O.}\ \bibnamefont {{Gough}}}, \bibinfo {author}
  {\bibfnamefont {G.~R.}\ \bibnamefont {{Isaak}}},\ and\ \bibinfo {author}
  {\bibfnamefont {R.}~\bibnamefont {{New}}},\ }\bibfield  {title} {\bibinfo
  {title} {{On model predictions of the power spectral density of radial solar
  p modes}},\ }\href {https://doi.org/10.1111/j.1365-2966.2005.09041.x}
  {\bibfield  {journal} {\bibinfo  {journal} {\mnras}\ }\textbf {\bibinfo
  {volume} {360}},\ \bibinfo {pages} {859} (\bibinfo {year}
  {2005})}\BibitemShut {NoStop}%
\bibitem [{\citenamefont {Aerts}\ \emph {et~al.}(2010)\citenamefont {Aerts},
  \citenamefont {Christensen-Dalsgaard},\ and\ \citenamefont
  {Kurtz}}]{aerts2010asteroseismology}%
  \BibitemOpen
  \bibfield  {author} {\bibinfo {author} {\bibfnamefont {C.}~\bibnamefont
  {Aerts}}, \bibinfo {author} {\bibfnamefont {J.}~\bibnamefont
  {Christensen-Dalsgaard}},\ and\ \bibinfo {author} {\bibfnamefont
  {D.}~\bibnamefont {Kurtz}},\ }\href
  {https://books.google.com/books?id=N8pswDrdSyUC} {\emph {\bibinfo {title}
  {Asteroseismology}}},\ Astronomy and Astrophysics Library\ (\bibinfo
  {publisher} {Springer Netherlands},\ \bibinfo {year} {2010})\BibitemShut
  {NoStop}%
\bibitem [{\citenamefont {Lopes}\ and\ \citenamefont
  {Silk}(2014)}]{Lopes:2014dba}%
  \BibitemOpen
  \bibfield  {author} {\bibinfo {author} {\bibfnamefont {I.}~\bibnamefont
  {Lopes}}\ and\ \bibinfo {author} {\bibfnamefont {J.}~\bibnamefont {Silk}},\
  }\bibfield  {title} {\bibinfo {title} {{Helioseismology and Asteroseismology:
  Looking for Gravitational Waves in acoustic oscillations}},\ }\href
  {https://doi.org/10.1088/0004-637X/794/1/32} {\bibfield  {journal} {\bibinfo
  {journal} {Astrophys. J.}\ }\textbf {\bibinfo {volume} {794}},\ \bibinfo
  {pages} {32} (\bibinfo {year} {2014})},\ \Eprint
  {https://arxiv.org/abs/1405.0292} {arXiv:1405.0292 [astro-ph.CO]}
  \BibitemShut {NoStop}%
\bibitem [{\citenamefont {Lopes}\ and\ \citenamefont
  {Silk}(2015)}]{Lopes:2015pca}%
  \BibitemOpen
  \bibfield  {author} {\bibinfo {author} {\bibfnamefont {I.}~\bibnamefont
  {Lopes}}\ and\ \bibinfo {author} {\bibfnamefont {J.}~\bibnamefont {Silk}},\
  }\bibfield  {title} {\bibinfo {title} {{Nearby Stars as Gravitational Wave
  Detectors}},\ }\href {https://doi.org/10.1088/0004-637X/807/2/135} {\bibfield
   {journal} {\bibinfo  {journal} {Astrophys. J.}\ }\textbf {\bibinfo {volume}
  {807}},\ \bibinfo {pages} {135} (\bibinfo {year} {2015})},\ \Eprint
  {https://arxiv.org/abs/1507.03212} {arXiv:1507.03212 [astro-ph.SR]}
  \BibitemShut {NoStop}%
\bibitem [{\citenamefont {{Goldreich}}\ and\ \citenamefont
  {{Keeley}}(1977)}]{1977ApJ...212..243G}%
  \BibitemOpen
  \bibfield  {author} {\bibinfo {author} {\bibfnamefont {P.}~\bibnamefont
  {{Goldreich}}}\ and\ \bibinfo {author} {\bibfnamefont {D.~A.}\ \bibnamefont
  {{Keeley}}},\ }\bibfield  {title} {\bibinfo {title} {{Solar seismology. II.
  The stochastic excitation of the solar p-modes by turbulent convection.}},\
  }\href {https://doi.org/10.1086/155043} {\bibfield  {journal} {\bibinfo
  {journal} {\apj}\ }\textbf {\bibinfo {volume} {212}},\ \bibinfo {pages} {243}
  (\bibinfo {year} {1977})}\BibitemShut {NoStop}%
\bibitem [{\citenamefont {Chaplin}\ \emph {et~al.}(2009)\citenamefont
  {Chaplin}, \citenamefont {Houdek}, \citenamefont {Elsworth}, \citenamefont
  {New}, \citenamefont {Bedding},\ and\ \citenamefont
  {Kjeldsen}}]{Chaplin:2008af}%
  \BibitemOpen
  \bibfield  {author} {\bibinfo {author} {\bibfnamefont {W.~J.}\ \bibnamefont
  {Chaplin}}, \bibinfo {author} {\bibfnamefont {G.}~\bibnamefont {Houdek}},
  \bibinfo {author} {\bibfnamefont {Y.}~\bibnamefont {Elsworth}}, \bibinfo
  {author} {\bibfnamefont {R.}~\bibnamefont {New}}, \bibinfo {author}
  {\bibfnamefont {T.~R.}\ \bibnamefont {Bedding}},\ and\ \bibinfo {author}
  {\bibfnamefont {H.}~\bibnamefont {Kjeldsen}},\ }\bibfield  {title} {\bibinfo
  {title} {{Excitation and damping of p-mode oscillations of alpha Cen B}},\
  }\href {https://doi.org/10.1088/0004-637X/692/1/531} {\bibfield  {journal}
  {\bibinfo  {journal} {Astrophys. J.}\ }\textbf {\bibinfo {volume} {692}},\
  \bibinfo {pages} {531} (\bibinfo {year} {2009})},\ \Eprint
  {https://arxiv.org/abs/0810.5022} {arXiv:0810.5022 [astro-ph]} \BibitemShut
  {NoStop}%
\bibitem [{\citenamefont {{Houdek}}\ \emph {et~al.}(2019)\citenamefont
  {{Houdek}}, \citenamefont {{Lund}}, \citenamefont {{Trampedach}},
  \citenamefont {{Christensen-Dalsgaard}}, \citenamefont {{Handberg}},\ and\
  \citenamefont {{Appourchaux}}}]{2019MNRAS.487..595H}%
  \BibitemOpen
  \bibfield  {author} {\bibinfo {author} {\bibfnamefont {G.}~\bibnamefont
  {{Houdek}}}, \bibinfo {author} {\bibfnamefont {M.~N.}\ \bibnamefont
  {{Lund}}}, \bibinfo {author} {\bibfnamefont {R.}~\bibnamefont
  {{Trampedach}}}, \bibinfo {author} {\bibfnamefont {J.}~\bibnamefont
  {{Christensen-Dalsgaard}}}, \bibinfo {author} {\bibfnamefont
  {R.}~\bibnamefont {{Handberg}}},\ and\ \bibinfo {author} {\bibfnamefont
  {T.}~\bibnamefont {{Appourchaux}}},\ }\bibfield  {title} {\bibinfo {title}
  {{Damping rates and frequency corrections of Kepler LEGACY stars}},\ }\href
  {https://doi.org/10.1093/mnras/stz1211} {\bibfield  {journal} {\bibinfo
  {journal} {\mnras}\ }\textbf {\bibinfo {volume} {487}},\ \bibinfo {pages}
  {595} (\bibinfo {year} {2019})},\ \Eprint {https://arxiv.org/abs/1904.13170}
  {arXiv:1904.13170 [astro-ph.SR]} \BibitemShut {NoStop}%
\bibitem [{\citenamefont {{Houdek}}(2006)}]{2006ESASP.624E..28H}%
  \BibitemOpen
  \bibfield  {author} {\bibinfo {author} {\bibfnamefont {G.}~\bibnamefont
  {{Houdek}}},\ }\bibfield  {title} {\bibinfo {title} {{Stochastic excitation
  and damping of solar-like oscillations}},\ }in\ \href@noop {} {\emph
  {\bibinfo {booktitle} {Proceedings of SOHO 18/GONG 2006/HELAS I, Beyond the
  spherical Sun}}},\ \bibinfo {series} {ESA Special Publication}, Vol.\
  \bibinfo {volume} {624},\ \bibinfo {editor} {edited by\ \bibinfo {editor}
  {\bibfnamefont {K.}~\bibnamefont {{Fletcher}}}\ and\ \bibinfo {editor}
  {\bibfnamefont {M.}~\bibnamefont {{Thompson}}}}\ (\bibinfo {year} {2006})\
  p.~\bibinfo {pages} {28}\BibitemShut {NoStop}%
\bibitem [{\citenamefont
  {Christensen-Dalsgaard}(2003)}]{Christensen-Dalsgaard:2002ney}%
  \BibitemOpen
  \bibfield  {author} {\bibinfo {author} {\bibfnamefont {J.}~\bibnamefont
  {Christensen-Dalsgaard}},\ }\bibfield  {title} {\bibinfo {title}
  {{Helioseismology}},\ }\href {https://doi.org/10.1103/RevModPhys.74.1073}
  {\bibfield  {journal} {\bibinfo  {journal} {Rev. Mod. Phys.}\ }\textbf
  {\bibinfo {volume} {74}},\ \bibinfo {pages} {1073} (\bibinfo {year}
  {2003})},\ \Eprint {https://arxiv.org/abs/astro-ph/0207403}
  {arXiv:astro-ph/0207403} \BibitemShut {NoStop}%
\bibitem [{\citenamefont {Kolmogorov}\ \emph {et~al.}(1991)\citenamefont
  {Kolmogorov}, \citenamefont {Levin}, \citenamefont {Hunt}, \citenamefont
  {Phillips},\ and\ \citenamefont {Williams}}]{Kolmogorov1890}%
  \BibitemOpen
  \bibfield  {author} {\bibinfo {author} {\bibfnamefont {A.~N.}\ \bibnamefont
  {Kolmogorov}}, \bibinfo {author} {\bibfnamefont {V.}~\bibnamefont {Levin}},
  \bibinfo {author} {\bibfnamefont {J.~C.~R.}\ \bibnamefont {Hunt}}, \bibinfo
  {author} {\bibfnamefont {O.~M.}\ \bibnamefont {Phillips}},\ and\ \bibinfo
  {author} {\bibfnamefont {D.}~\bibnamefont {Williams}},\ }\bibfield  {title}
  {\bibinfo {title} {The local structure of turbulence in incompressible
  viscous fluid for very large reynolds numbers},\ }\href
  {https://doi.org/10.1098/rspa.1991.0075} {\bibfield  {journal} {\bibinfo
  {journal} {Proceedings of the Royal Society of London. Series A: Mathematical
  and Physical Sciences}\ }\textbf {\bibinfo {volume} {434}},\ \bibinfo {pages}
  {9} (\bibinfo {year} {1991})},\ \Eprint
  {https://arxiv.org/abs/https://royalsocietypublishing.org/doi/pdf/10.1098/rspa.1991.0075}
  {https://royalsocietypublishing.org/doi/pdf/10.1098/rspa.1991.0075}
  \BibitemShut {NoStop}%
\bibitem [{\citenamefont {Samadi}\ and\ \citenamefont
  {Goupil}(2001)}]{Samadi:2001nc}%
  \BibitemOpen
  \bibfield  {author} {\bibinfo {author} {\bibfnamefont {R.}~\bibnamefont
  {Samadi}}\ and\ \bibinfo {author} {\bibfnamefont {M.-J.}\ \bibnamefont
  {Goupil}},\ }\bibfield  {title} {\bibinfo {title} {{Excitation of stellar
  p-modes by turbulent convection: 1. Theoretical formulation}},\ }\href
  {https://doi.org/10.1051/0004-6361:20010212} {\bibfield  {journal} {\bibinfo
  {journal} {Astron. Astrophys.}\ }\textbf {\bibinfo {volume} {370}},\ \bibinfo
  {pages} {136} (\bibinfo {year} {2001})},\ \Eprint
  {https://arxiv.org/abs/astro-ph/0101109} {arXiv:astro-ph/0101109}
  \BibitemShut {NoStop}%
\bibitem [{\citenamefont {{Houdek}}(1996)}]{GoudekPhD}%
  \BibitemOpen
  \bibfield  {author} {\bibinfo {author} {\bibfnamefont {G.}~\bibnamefont
  {{Houdek}}},\ }\emph {\bibinfo {title} {{Pulsation of Solar-type stars}}},\
  \href@noop {} {Ph.D. thesis},\ \bibinfo  {school} {-} (\bibinfo {year}
  {1996})\BibitemShut {NoStop}%
\bibitem [{\citenamefont {{Libbrecht}}(1988)}]{Libbrecht1998}%
  \BibitemOpen
  \bibfield  {author} {\bibinfo {author} {\bibfnamefont {K.~G.}\ \bibnamefont
  {{Libbrecht}}},\ }\bibfield  {title} {\bibinfo {title} {{Solar p-Mode
  Phenomenology}},\ }\href {https://doi.org/10.1086/166855} {\bibfield
  {journal} {\bibinfo  {journal} {\apj}\ }\textbf {\bibinfo {volume} {334}},\
  \bibinfo {pages} {510} (\bibinfo {year} {1988})}\BibitemShut {NoStop}%
\bibitem [{\citenamefont {{Paxton}}\ \emph {et~al.}(2011)\citenamefont
  {{Paxton}}, \citenamefont {{Bildsten}}, \citenamefont {{Dotter}},
  \citenamefont {{Herwig}}, \citenamefont {{Lesaffre}},\ and\ \citenamefont
  {{Timmes}}}]{Paxton2011}%
  \BibitemOpen
  \bibfield  {author} {\bibinfo {author} {\bibfnamefont {B.}~\bibnamefont
  {{Paxton}}}, \bibinfo {author} {\bibfnamefont {L.}~\bibnamefont
  {{Bildsten}}}, \bibinfo {author} {\bibfnamefont {A.}~\bibnamefont
  {{Dotter}}}, \bibinfo {author} {\bibfnamefont {F.}~\bibnamefont {{Herwig}}},
  \bibinfo {author} {\bibfnamefont {P.}~\bibnamefont {{Lesaffre}}},\ and\
  \bibinfo {author} {\bibfnamefont {F.}~\bibnamefont {{Timmes}}},\ }\bibfield
  {title} {\bibinfo {title} {{Modules for Experiments in Stellar Astrophysics
  (MESA)}},\ }\href {https://doi.org/10.1088/0067-0049/192/1/3} {\bibfield
  {journal} {\bibinfo  {journal} {The Astrophysical Journal Supplement Series}\
  }\textbf {\bibinfo {volume} {192}},\ \bibinfo {eid} {3} (\bibinfo {year}
  {2011})},\ \Eprint {https://arxiv.org/abs/1009.1622} {arXiv:1009.1622
  [astro-ph.SR]} \BibitemShut {NoStop}%
\bibitem [{\citenamefont {{Paxton}}\ \emph {et~al.}(2013)\citenamefont
  {{Paxton}}, \citenamefont {{Cantiello}}, \citenamefont {{Arras}},
  \citenamefont {{Bildsten}}, \citenamefont {{Brown}}, \citenamefont
  {{Dotter}}, \citenamefont {{Mankovich}}, \citenamefont {{Montgomery}},
  \citenamefont {{Stello}}, \citenamefont {{Timmes}},\ and\ \citenamefont
  {{Townsend}}}]{Paxton2013}%
  \BibitemOpen
  \bibfield  {author} {\bibinfo {author} {\bibfnamefont {B.}~\bibnamefont
  {{Paxton}}}, \bibinfo {author} {\bibfnamefont {M.}~\bibnamefont
  {{Cantiello}}}, \bibinfo {author} {\bibfnamefont {P.}~\bibnamefont
  {{Arras}}}, \bibinfo {author} {\bibfnamefont {L.}~\bibnamefont {{Bildsten}}},
  \bibinfo {author} {\bibfnamefont {E.~F.}\ \bibnamefont {{Brown}}}, \bibinfo
  {author} {\bibfnamefont {A.}~\bibnamefont {{Dotter}}}, \bibinfo {author}
  {\bibfnamefont {C.}~\bibnamefont {{Mankovich}}}, \bibinfo {author}
  {\bibfnamefont {M.~H.}\ \bibnamefont {{Montgomery}}}, \bibinfo {author}
  {\bibfnamefont {D.}~\bibnamefont {{Stello}}}, \bibinfo {author}
  {\bibfnamefont {F.~X.}\ \bibnamefont {{Timmes}}},\ and\ \bibinfo {author}
  {\bibfnamefont {R.}~\bibnamefont {{Townsend}}},\ }\bibfield  {title}
  {\bibinfo {title} {{Modules for Experiments in Stellar Astrophysics (MESA):
  Planets, Oscillations, Rotation, and Massive Stars}},\ }\href
  {https://doi.org/10.1088/0067-0049/208/1/4} {\bibfield  {journal} {\bibinfo
  {journal} {The Astrophysical Journal Supplement Series}\ }\textbf {\bibinfo
  {volume} {208}},\ \bibinfo {eid} {4} (\bibinfo {year} {2013})},\ \Eprint
  {https://arxiv.org/abs/1301.0319} {arXiv:1301.0319 [astro-ph.SR]}
  \BibitemShut {NoStop}%
\bibitem [{\citenamefont {{Paxton}}\ \emph {et~al.}(2015)\citenamefont
  {{Paxton}}, \citenamefont {{Marchant}}, \citenamefont {{Schwab}},
  \citenamefont {{Bauer}}, \citenamefont {{Bildsten}}, \citenamefont
  {{Cantiello}}, \citenamefont {{Dessart}}, \citenamefont {{Farmer}},
  \citenamefont {{Hu}}, \citenamefont {{Langer}}, \citenamefont {{Townsend}},
  \citenamefont {{Townsley}},\ and\ \citenamefont {{Timmes}}}]{Paxton2015}%
  \BibitemOpen
  \bibfield  {author} {\bibinfo {author} {\bibfnamefont {B.}~\bibnamefont
  {{Paxton}}}, \bibinfo {author} {\bibfnamefont {P.}~\bibnamefont
  {{Marchant}}}, \bibinfo {author} {\bibfnamefont {J.}~\bibnamefont
  {{Schwab}}}, \bibinfo {author} {\bibfnamefont {E.~B.}\ \bibnamefont
  {{Bauer}}}, \bibinfo {author} {\bibfnamefont {L.}~\bibnamefont {{Bildsten}}},
  \bibinfo {author} {\bibfnamefont {M.}~\bibnamefont {{Cantiello}}}, \bibinfo
  {author} {\bibfnamefont {L.}~\bibnamefont {{Dessart}}}, \bibinfo {author}
  {\bibfnamefont {R.}~\bibnamefont {{Farmer}}}, \bibinfo {author}
  {\bibfnamefont {H.}~\bibnamefont {{Hu}}}, \bibinfo {author} {\bibfnamefont
  {N.}~\bibnamefont {{Langer}}}, \bibinfo {author} {\bibfnamefont {R.~H.~D.}\
  \bibnamefont {{Townsend}}}, \bibinfo {author} {\bibfnamefont {D.~M.}\
  \bibnamefont {{Townsley}}},\ and\ \bibinfo {author} {\bibfnamefont {F.~X.}\
  \bibnamefont {{Timmes}}},\ }\bibfield  {title} {\bibinfo {title} {{Modules
  for Experiments in Stellar Astrophysics (MESA): Binaries, Pulsations, and
  Explosions}},\ }\href {https://doi.org/10.1088/0067-0049/220/1/15} {\bibfield
   {journal} {\bibinfo  {journal} {The Astrophysical Journal Supplement
  Series}\ }\textbf {\bibinfo {volume} {220}},\ \bibinfo {eid} {15} (\bibinfo
  {year} {2015})},\ \Eprint {https://arxiv.org/abs/1506.03146}
  {arXiv:1506.03146 [astro-ph.SR]} \BibitemShut {NoStop}%
\bibitem [{\citenamefont {{Paxton}}\ \emph {et~al.}(2018)\citenamefont
  {{Paxton}}, \citenamefont {{Schwab}}, \citenamefont {{Bauer}}, \citenamefont
  {{Bildsten}}, \citenamefont {{Blinnikov}}, \citenamefont {{Duffell}},
  \citenamefont {{Farmer}}, \citenamefont {{Goldberg}}, \citenamefont
  {{Marchant}}, \citenamefont {{Sorokina}}, \citenamefont {{Thoul}},
  \citenamefont {{Townsend}},\ and\ \citenamefont {{Timmes}}}]{Paxton2018}%
  \BibitemOpen
  \bibfield  {author} {\bibinfo {author} {\bibfnamefont {B.}~\bibnamefont
  {{Paxton}}}, \bibinfo {author} {\bibfnamefont {J.}~\bibnamefont {{Schwab}}},
  \bibinfo {author} {\bibfnamefont {E.~B.}\ \bibnamefont {{Bauer}}}, \bibinfo
  {author} {\bibfnamefont {L.}~\bibnamefont {{Bildsten}}}, \bibinfo {author}
  {\bibfnamefont {S.}~\bibnamefont {{Blinnikov}}}, \bibinfo {author}
  {\bibfnamefont {P.}~\bibnamefont {{Duffell}}}, \bibinfo {author}
  {\bibfnamefont {R.}~\bibnamefont {{Farmer}}}, \bibinfo {author}
  {\bibfnamefont {J.~A.}\ \bibnamefont {{Goldberg}}}, \bibinfo {author}
  {\bibfnamefont {P.}~\bibnamefont {{Marchant}}}, \bibinfo {author}
  {\bibfnamefont {E.}~\bibnamefont {{Sorokina}}}, \bibinfo {author}
  {\bibfnamefont {A.}~\bibnamefont {{Thoul}}}, \bibinfo {author} {\bibfnamefont
  {R.~H.~D.}\ \bibnamefont {{Townsend}}},\ and\ \bibinfo {author}
  {\bibfnamefont {F.~X.}\ \bibnamefont {{Timmes}}},\ }\bibfield  {title}
  {\bibinfo {title} {{Modules for Experiments in Stellar Astrophysics (MESA):
  Convective Boundaries, Element Diffusion, and Massive Star Explosions}},\
  }\href {https://doi.org/10.3847/1538-4365/aaa5a8} {\bibfield  {journal}
  {\bibinfo  {journal} {The Astrophysical Journal Supplement Series}\ }\textbf
  {\bibinfo {volume} {234}},\ \bibinfo {eid} {34} (\bibinfo {year} {2018})},\
  \Eprint {https://arxiv.org/abs/1710.08424} {arXiv:1710.08424 [astro-ph.SR]}
  \BibitemShut {NoStop}%
\bibitem [{\citenamefont {{Paxton}}\ \emph {et~al.}(2019)\citenamefont
  {{Paxton}}, \citenamefont {{Smolec}}, \citenamefont {{Gautschy}},
  \citenamefont {{Bildsten}}, \citenamefont {{Cantiello}}, \citenamefont
  {{Dotter}}, \citenamefont {{Farmer}}, \citenamefont {{Goldberg}},
  \citenamefont {{Jermyn}}, \citenamefont {{Kanbur}}, \citenamefont
  {{Marchant}}, \citenamefont {{Schwab}}, \citenamefont {{Thoul}},
  \citenamefont {{Townsend}}, \citenamefont {{Wolf}}, \citenamefont {{Zhang}},\
  and\ \citenamefont {{Timmes}}}]{Paxton2019}%
  \BibitemOpen
  \bibfield  {author} {\bibinfo {author} {\bibfnamefont {B.}~\bibnamefont
  {{Paxton}}}, \bibinfo {author} {\bibfnamefont {R.}~\bibnamefont {{Smolec}}},
  \bibinfo {author} {\bibfnamefont {A.}~\bibnamefont {{Gautschy}}}, \bibinfo
  {author} {\bibfnamefont {L.}~\bibnamefont {{Bildsten}}}, \bibinfo {author}
  {\bibfnamefont {M.}~\bibnamefont {{Cantiello}}}, \bibinfo {author}
  {\bibfnamefont {A.}~\bibnamefont {{Dotter}}}, \bibinfo {author}
  {\bibfnamefont {R.}~\bibnamefont {{Farmer}}}, \bibinfo {author}
  {\bibfnamefont {J.~A.}\ \bibnamefont {{Goldberg}}}, \bibinfo {author}
  {\bibfnamefont {A.~S.}\ \bibnamefont {{Jermyn}}}, \bibinfo {author}
  {\bibfnamefont {S.~M.}\ \bibnamefont {{Kanbur}}}, \bibinfo {author}
  {\bibfnamefont {P.}~\bibnamefont {{Marchant}}}, \bibinfo {author}
  {\bibfnamefont {J.}~\bibnamefont {{Schwab}}}, \bibinfo {author}
  {\bibfnamefont {A.}~\bibnamefont {{Thoul}}}, \bibinfo {author} {\bibfnamefont
  {R.~H.~D.}\ \bibnamefont {{Townsend}}}, \bibinfo {author} {\bibfnamefont
  {W.~M.}\ \bibnamefont {{Wolf}}}, \bibinfo {author} {\bibfnamefont
  {M.}~\bibnamefont {{Zhang}}},\ and\ \bibinfo {author} {\bibfnamefont {F.~X.}\
  \bibnamefont {{Timmes}}},\ }\bibfield  {title} {\bibinfo {title} {{Modules
  for Experiments in Stellar Astrophysics (MESA): Pulsating Variable Stars,
  Rotation, Convective Boundaries, and Energy Conservation}},\ }\href@noop {}
  {\bibfield  {journal} {\bibinfo  {journal} {arXiv e-prints}\ } (\bibinfo
  {year} {2019})},\ \Eprint {https://arxiv.org/abs/1903.01426}
  {arXiv:1903.01426 [astro-ph.SR]} \BibitemShut {NoStop}%
\bibitem [{\citenamefont {{Townsend}}\ and\ \citenamefont
  {{Teitler}}(2013)}]{2013MNRAS.435.3406T}%
  \BibitemOpen
  \bibfield  {author} {\bibinfo {author} {\bibfnamefont {R.~H.~D.}\
  \bibnamefont {{Townsend}}}\ and\ \bibinfo {author} {\bibfnamefont {S.~A.}\
  \bibnamefont {{Teitler}}},\ }\bibfield  {title} {\bibinfo {title} {{GYRE: an
  open-source stellar oscillation code based on a new Magnus Multiple Shooting
  scheme}},\ }\href {https://doi.org/10.1093/mnras/stt1533} {\bibfield
  {journal} {\bibinfo  {journal} {\mnras}\ }\textbf {\bibinfo {volume} {435}},\
  \bibinfo {pages} {3406} (\bibinfo {year} {2013})},\ \Eprint
  {https://arxiv.org/abs/1308.2965} {arXiv:1308.2965 [astro-ph.SR]}
  \BibitemShut {NoStop}%
\bibitem [{\citenamefont {Saltas}\ and\ \citenamefont
  {Christensen-Dalsgaard}(2022)}]{Saltas:2022ybg}%
  \BibitemOpen
  \bibfield  {author} {\bibinfo {author} {\bibfnamefont {I.~D.}\ \bibnamefont
  {Saltas}}\ and\ \bibinfo {author} {\bibfnamefont {J.}~\bibnamefont
  {Christensen-Dalsgaard}},\ }\bibfield  {title} {\bibinfo {title} {{Searching
  for dark energy with the Sun}},\ }\href
  {https://doi.org/10.1051/0004-6361/202244176} {\bibfield  {journal} {\bibinfo
   {journal} {Astron. Astrophys.}\ }\textbf {\bibinfo {volume} {667}},\
  \bibinfo {pages} {A115} (\bibinfo {year} {2022})},\ \Eprint
  {https://arxiv.org/abs/2205.14134} {arXiv:2205.14134 [astro-ph.SR]}
  \BibitemShut {NoStop}%
\bibitem [{\citenamefont {{Grevesse}}\ and\ \citenamefont
  {{Sauval}}(1998)}]{GS98}%
  \BibitemOpen
  \bibfield  {author} {\bibinfo {author} {\bibfnamefont {N.}~\bibnamefont
  {{Grevesse}}}\ and\ \bibinfo {author} {\bibfnamefont {A.~J.}\ \bibnamefont
  {{Sauval}}},\ }\bibfield  {title} {\bibinfo {title} {{Standard Solar
  Composition}},\ }\href {https://doi.org/10.1023/A:1005161325181} {\bibfield
  {journal} {\bibinfo  {journal} {\ssr}\ }\textbf {\bibinfo {volume} {85}},\
  \bibinfo {pages} {161} (\bibinfo {year} {1998})}\BibitemShut {NoStop}%
\bibitem [{\citenamefont {{Rogers}}\ and\ \citenamefont
  {{Nayfonov}}(2002)}]{Rogers2002}%
  \BibitemOpen
  \bibfield  {author} {\bibinfo {author} {\bibfnamefont {F.~J.}\ \bibnamefont
  {{Rogers}}}\ and\ \bibinfo {author} {\bibfnamefont {A.}~\bibnamefont
  {{Nayfonov}}},\ }\bibfield  {title} {\bibinfo {title} {{Updated and Expanded
  OPAL Equation-of-State Tables: Implications for Helioseismology}},\ }\href
  {https://doi.org/10.1086/341894} {\bibfield  {journal} {\bibinfo  {journal}
  {The Astrophysical Journal}\ }\textbf {\bibinfo {volume} {576}},\ \bibinfo
  {pages} {1064} (\bibinfo {year} {2002})}\BibitemShut {NoStop}%
\bibitem [{\citenamefont {{Seaton}}(2005)}]{Seaton2005}%
  \BibitemOpen
  \bibfield  {author} {\bibinfo {author} {\bibfnamefont {M.~J.}\ \bibnamefont
  {{Seaton}}},\ }\bibfield  {title} {\bibinfo {title} {{Opacity Project data on
  CD for mean opacities and radiative accelerations}},\ }\href
  {https://doi.org/10.1111/j.1365-2966.2005.00019.x} {\bibfield  {journal}
  {\bibinfo  {journal} {\mnras}\ }\textbf {\bibinfo {volume} {362}},\ \bibinfo
  {pages} {L1} (\bibinfo {year} {2005})},\ \Eprint
  {https://arxiv.org/abs/astro-ph/0411010} {arXiv:astro-ph/0411010 [astro-ph]}
  \BibitemShut {NoStop}%
\bibitem [{\citenamefont {{Sofue}}(2020)}]{2020Galax...8...37S}%
  \BibitemOpen
  \bibfield  {author} {\bibinfo {author} {\bibfnamefont {Y.}~\bibnamefont
  {{Sofue}}},\ }\bibfield  {title} {\bibinfo {title} {{Rotation Curve of the
  Milky Way and the Dark Matter Density}},\ }\href
  {https://doi.org/10.3390/galaxies8020037} {\bibfield  {journal} {\bibinfo
  {journal} {Galaxies}\ }\textbf {\bibinfo {volume} {8}},\ \bibinfo {pages}
  {37} (\bibinfo {year} {2020})},\ \Eprint {https://arxiv.org/abs/2004.11688}
  {arXiv:2004.11688 [astro-ph.GA]} \BibitemShut {NoStop}%
\bibitem [{\citenamefont {{Belkacem, K.}}\ \emph {et~al.}(2012)\citenamefont
  {{Belkacem, K.}}, \citenamefont {{Dupret, M. A.}}, \citenamefont {{Baudin,
  F.}}, \citenamefont {{Appourchaux, T.}}, \citenamefont {{Marques, J. P.}},\
  and\ \citenamefont {{Samadi, R.}}}]{Belkacem2012}%
  \BibitemOpen
  \bibfield  {author} {\bibinfo {author} {\bibnamefont {{Belkacem, K.}}},
  \bibinfo {author} {\bibnamefont {{Dupret, M. A.}}}, \bibinfo {author}
  {\bibnamefont {{Baudin, F.}}}, \bibinfo {author} {\bibnamefont {{Appourchaux,
  T.}}}, \bibinfo {author} {\bibnamefont {{Marques, J. P.}}},\ and\ \bibinfo
  {author} {\bibnamefont {{Samadi, R.}}},\ }\bibfield  {title} {\bibinfo
  {title} {Damping rates of solar-like oscillations across the hr diagram -
  theoretical calculations confronted to corot and kepler observations},\
  }\href {https://doi.org/10.1051/0004-6361/201218890} {\bibfield  {journal}
  {\bibinfo  {journal} {A\&A}\ }\textbf {\bibinfo {volume} {540}},\ \bibinfo
  {pages} {L7} (\bibinfo {year} {2012})}\BibitemShut {NoStop}%
\bibitem [{\citenamefont {{Huber}}\ \emph {et~al.}(2010)\citenamefont
  {{Huber}}, \citenamefont {{Bedding}}, \citenamefont {{Stello}}, \citenamefont
  {{Mosser}}, \citenamefont {{Mathur}}, \citenamefont {{Kallinger}},
  \citenamefont {{Hekker}}, \citenamefont {{Elsworth}}, \citenamefont
  {{Buzasi}}, \citenamefont {{De Ridder}}, \citenamefont {{Gilliland}},
  \citenamefont {{Kjeldsen}}, \citenamefont {{Chaplin}}, \citenamefont
  {{Garc{\'\i}a}}, \citenamefont {{Hale}}, \citenamefont {{Preston}},
  \citenamefont {{White}}, \citenamefont {{Borucki}}, \citenamefont
  {{Christensen-Dalsgaard}}, \citenamefont {{Clarke}}, \citenamefont
  {{Jenkins}},\ and\ \citenamefont {{Koch}}}]{2010ApJ...723.1607H}%
  \BibitemOpen
  \bibfield  {author} {\bibinfo {author} {\bibfnamefont {D.}~\bibnamefont
  {{Huber}}}, \bibinfo {author} {\bibfnamefont {T.~R.}\ \bibnamefont
  {{Bedding}}}, \bibinfo {author} {\bibfnamefont {D.}~\bibnamefont {{Stello}}},
  \bibinfo {author} {\bibfnamefont {B.}~\bibnamefont {{Mosser}}}, \bibinfo
  {author} {\bibfnamefont {S.}~\bibnamefont {{Mathur}}}, \bibinfo {author}
  {\bibfnamefont {T.}~\bibnamefont {{Kallinger}}}, \bibinfo {author}
  {\bibfnamefont {S.}~\bibnamefont {{Hekker}}}, \bibinfo {author}
  {\bibfnamefont {Y.~P.}\ \bibnamefont {{Elsworth}}}, \bibinfo {author}
  {\bibfnamefont {D.~L.}\ \bibnamefont {{Buzasi}}}, \bibinfo {author}
  {\bibfnamefont {J.}~\bibnamefont {{De Ridder}}}, \bibinfo {author}
  {\bibfnamefont {R.~L.}\ \bibnamefont {{Gilliland}}}, \bibinfo {author}
  {\bibfnamefont {H.}~\bibnamefont {{Kjeldsen}}}, \bibinfo {author}
  {\bibfnamefont {W.~J.}\ \bibnamefont {{Chaplin}}}, \bibinfo {author}
  {\bibfnamefont {R.~A.}\ \bibnamefont {{Garc{\'\i}a}}}, \bibinfo {author}
  {\bibfnamefont {S.~J.}\ \bibnamefont {{Hale}}}, \bibinfo {author}
  {\bibfnamefont {H.~L.}\ \bibnamefont {{Preston}}}, \bibinfo {author}
  {\bibfnamefont {T.~R.}\ \bibnamefont {{White}}}, \bibinfo {author}
  {\bibfnamefont {W.~J.}\ \bibnamefont {{Borucki}}}, \bibinfo {author}
  {\bibfnamefont {J.}~\bibnamefont {{Christensen-Dalsgaard}}}, \bibinfo
  {author} {\bibfnamefont {B.~D.}\ \bibnamefont {{Clarke}}}, \bibinfo {author}
  {\bibfnamefont {J.~M.}\ \bibnamefont {{Jenkins}}},\ and\ \bibinfo {author}
  {\bibfnamefont {D.}~\bibnamefont {{Koch}}},\ }\bibfield  {title} {\bibinfo
  {title} {{Asteroseismology of Red Giants from the First Four Months of Kepler
  Data: Global Oscillation Parameters for 800 Stars}},\ }\href
  {https://doi.org/10.1088/0004-637X/723/2/1607} {\bibfield  {journal}
  {\bibinfo  {journal} {\apj}\ }\textbf {\bibinfo {volume} {723}},\ \bibinfo
  {pages} {1607} (\bibinfo {year} {2010})},\ \Eprint
  {https://arxiv.org/abs/1010.4566} {arXiv:1010.4566 [astro-ph.SR]}
  \BibitemShut {NoStop}%
\bibitem [{\citenamefont {{Yu}}\ \emph {et~al.}(2018)\citenamefont {{Yu}},
  \citenamefont {{Huber}}, \citenamefont {{Bedding}}, \citenamefont {{Stello}},
  \citenamefont {{Hon}}, \citenamefont {{Murphy}},\ and\ \citenamefont
  {{Khanna}}}]{2018ApJS..236...42Y}%
  \BibitemOpen
  \bibfield  {author} {\bibinfo {author} {\bibfnamefont {J.}~\bibnamefont
  {{Yu}}}, \bibinfo {author} {\bibfnamefont {D.}~\bibnamefont {{Huber}}},
  \bibinfo {author} {\bibfnamefont {T.~R.}\ \bibnamefont {{Bedding}}}, \bibinfo
  {author} {\bibfnamefont {D.}~\bibnamefont {{Stello}}}, \bibinfo {author}
  {\bibfnamefont {M.}~\bibnamefont {{Hon}}}, \bibinfo {author} {\bibfnamefont
  {S.~J.}\ \bibnamefont {{Murphy}}},\ and\ \bibinfo {author} {\bibfnamefont
  {S.}~\bibnamefont {{Khanna}}},\ }\bibfield  {title} {\bibinfo {title}
  {{Asteroseismology of 16,000 Kepler Red Giants: Global Oscillation
  Parameters, Masses, and Radii}},\ }\href
  {https://doi.org/10.3847/1538-4365/aaaf74} {\bibfield  {journal} {\bibinfo
  {journal} {\apjs}\ }\textbf {\bibinfo {volume} {236}},\ \bibinfo {eid} {42}
  (\bibinfo {year} {2018})},\ \Eprint {https://arxiv.org/abs/1802.04455}
  {arXiv:1802.04455 [astro-ph.SR]} \BibitemShut {NoStop}%
\bibitem [{\citenamefont {{Hon}}\ \emph {et~al.}(2019)\citenamefont {{Hon}},
  \citenamefont {{Stello}}, \citenamefont {{Garc{\'\i}a}}, \citenamefont
  {{Mathur}}, \citenamefont {{Sharma}}, \citenamefont {{Colman}},\ and\
  \citenamefont {{Bugnet}}}]{2019MNRAS.485.5616H}%
  \BibitemOpen
  \bibfield  {author} {\bibinfo {author} {\bibfnamefont {M.}~\bibnamefont
  {{Hon}}}, \bibinfo {author} {\bibfnamefont {D.}~\bibnamefont {{Stello}}},
  \bibinfo {author} {\bibfnamefont {R.~A.}\ \bibnamefont {{Garc{\'\i}a}}},
  \bibinfo {author} {\bibfnamefont {S.}~\bibnamefont {{Mathur}}}, \bibinfo
  {author} {\bibfnamefont {S.}~\bibnamefont {{Sharma}}}, \bibinfo {author}
  {\bibfnamefont {I.~L.}\ \bibnamefont {{Colman}}},\ and\ \bibinfo {author}
  {\bibfnamefont {L.}~\bibnamefont {{Bugnet}}},\ }\bibfield  {title} {\bibinfo
  {title} {{A search for red giant solar-like oscillations in all Kepler
  data}},\ }\href {https://doi.org/10.1093/mnras/stz622} {\bibfield  {journal}
  {\bibinfo  {journal} {\mnras}\ }\textbf {\bibinfo {volume} {485}},\ \bibinfo
  {pages} {5616} (\bibinfo {year} {2019})},\ \Eprint
  {https://arxiv.org/abs/1903.00115} {arXiv:1903.00115 [astro-ph.SR]}
  \BibitemShut {NoStop}%
\bibitem [{\citenamefont {{Beck}}\ \emph {et~al.}(2022)\citenamefont {{Beck}},
  \citenamefont {{Mathur}}, \citenamefont {{Hambleton}}, \citenamefont
  {{Garc{\'\i}a}}, \citenamefont {{Steinwender}}, \citenamefont {{Eisner}},
  \citenamefont {{do Nascimento}}, \citenamefont {{Gaulme}},\ and\
  \citenamefont {{Mathis}}}]{2022A&A...667A..31B}%
  \BibitemOpen
  \bibfield  {author} {\bibinfo {author} {\bibfnamefont {P.~G.}\ \bibnamefont
  {{Beck}}}, \bibinfo {author} {\bibfnamefont {S.}~\bibnamefont {{Mathur}}},
  \bibinfo {author} {\bibfnamefont {K.}~\bibnamefont {{Hambleton}}}, \bibinfo
  {author} {\bibfnamefont {R.~A.}\ \bibnamefont {{Garc{\'\i}a}}}, \bibinfo
  {author} {\bibfnamefont {L.}~\bibnamefont {{Steinwender}}}, \bibinfo {author}
  {\bibfnamefont {N.~L.}\ \bibnamefont {{Eisner}}}, \bibinfo {author}
  {\bibfnamefont {J.~D.}\ \bibnamefont {{do Nascimento}}}, \bibinfo {author}
  {\bibfnamefont {P.}~\bibnamefont {{Gaulme}}},\ and\ \bibinfo {author}
  {\bibfnamefont {S.}~\bibnamefont {{Mathis}}},\ }\bibfield  {title} {\bibinfo
  {title} {{99 oscillating red-giant stars in binary systems with NASA TESS and
  NASA Kepler identified from the SB9-Catalogue}},\ }\href
  {https://doi.org/10.1051/0004-6361/202143005} {\bibfield  {journal} {\bibinfo
   {journal} {\aap}\ }\textbf {\bibinfo {volume} {667}},\ \bibinfo {eid} {A31}
  (\bibinfo {year} {2022})},\ \Eprint {https://arxiv.org/abs/2202.02373}
  {arXiv:2202.02373 [astro-ph.SR]} \BibitemShut {NoStop}%
\bibitem [{\citenamefont {{Stello}}\ \emph {et~al.}(2022)\citenamefont
  {{Stello}}, \citenamefont {{Saunders}}, \citenamefont {{Grunblatt}},
  \citenamefont {{Hon}}, \citenamefont {{Reyes}}, \citenamefont {{Huber}},
  \citenamefont {{Bedding}}, \citenamefont {{Elsworth}}, \citenamefont
  {{Garc{\'\i}a}}, \citenamefont {{Hekker}}, \citenamefont {{Kallinger}},
  \citenamefont {{Mathur}}, \citenamefont {{Mosser}},\ and\ \citenamefont
  {{Pinsonneault}}}]{2022MNRAS.512.1677S}%
  \BibitemOpen
  \bibfield  {author} {\bibinfo {author} {\bibfnamefont {D.}~\bibnamefont
  {{Stello}}}, \bibinfo {author} {\bibfnamefont {N.}~\bibnamefont
  {{Saunders}}}, \bibinfo {author} {\bibfnamefont {S.}~\bibnamefont
  {{Grunblatt}}}, \bibinfo {author} {\bibfnamefont {M.}~\bibnamefont {{Hon}}},
  \bibinfo {author} {\bibfnamefont {C.}~\bibnamefont {{Reyes}}}, \bibinfo
  {author} {\bibfnamefont {D.}~\bibnamefont {{Huber}}}, \bibinfo {author}
  {\bibfnamefont {T.~R.}\ \bibnamefont {{Bedding}}}, \bibinfo {author}
  {\bibfnamefont {Y.}~\bibnamefont {{Elsworth}}}, \bibinfo {author}
  {\bibfnamefont {R.~A.}\ \bibnamefont {{Garc{\'\i}a}}}, \bibinfo {author}
  {\bibfnamefont {S.}~\bibnamefont {{Hekker}}}, \bibinfo {author}
  {\bibfnamefont {T.}~\bibnamefont {{Kallinger}}}, \bibinfo {author}
  {\bibfnamefont {S.}~\bibnamefont {{Mathur}}}, \bibinfo {author}
  {\bibfnamefont {B.}~\bibnamefont {{Mosser}}},\ and\ \bibinfo {author}
  {\bibfnamefont {M.~H.}\ \bibnamefont {{Pinsonneault}}},\ }\bibfield  {title}
  {\bibinfo {title} {{TESS asteroseismology of the Kepler red giants}},\ }\href
  {https://doi.org/10.1093/mnras/stac414} {\bibfield  {journal} {\bibinfo
  {journal} {\mnras}\ }\textbf {\bibinfo {volume} {512}},\ \bibinfo {pages}
  {1677} (\bibinfo {year} {2022})},\ \Eprint {https://arxiv.org/abs/2107.05831}
  {arXiv:2107.05831 [astro-ph.SR]} \BibitemShut {NoStop}%
\bibitem [{\citenamefont {{Hon}}\ \emph {et~al.}(2022)\citenamefont {{Hon}},
  \citenamefont {{Kuszlewicz}}, \citenamefont {{Huber}}, \citenamefont
  {{Stello}},\ and\ \citenamefont {{Reyes}}}]{2022AJ....164..135H}%
  \BibitemOpen
  \bibfield  {author} {\bibinfo {author} {\bibfnamefont {M.}~\bibnamefont
  {{Hon}}}, \bibinfo {author} {\bibfnamefont {J.~S.}\ \bibnamefont
  {{Kuszlewicz}}}, \bibinfo {author} {\bibfnamefont {D.}~\bibnamefont
  {{Huber}}}, \bibinfo {author} {\bibfnamefont {D.}~\bibnamefont {{Stello}}},\
  and\ \bibinfo {author} {\bibfnamefont {C.}~\bibnamefont {{Reyes}}},\
  }\bibfield  {title} {\bibinfo {title} {{HD-TESS: An Asteroseismic Catalog of
  Bright Red Giants within TESS Continuous Viewing Zones}},\ }\href
  {https://doi.org/10.3847/1538-3881/ac8931} {\bibfield  {journal} {\bibinfo
  {journal} {\aj}\ }\textbf {\bibinfo {volume} {164}},\ \bibinfo {eid} {135}
  (\bibinfo {year} {2022})},\ \Eprint {https://arxiv.org/abs/2208.06478}
  {arXiv:2208.06478 [astro-ph.SR]} \BibitemShut {NoStop}%
\bibitem [{\citenamefont {{Silva Aguirre}}\ \emph {et~al.}(2020)\citenamefont
  {{Silva Aguirre}}, \citenamefont {{Stello}}, \citenamefont {{Stokholm}},
  \citenamefont {{Mosumgaard}}, \citenamefont {{Ball}}, \citenamefont {{Basu}},
  \citenamefont {{Bossini}}, \citenamefont {{Bugnet}}, \citenamefont
  {{Buzasi}}, \citenamefont {{Campante}}, \citenamefont {{Carboneau}},
  \citenamefont {{Chaplin}}, \citenamefont {{Corsaro}}, \citenamefont
  {{Davies}}, \citenamefont {{Elsworth}}, \citenamefont {{Garc{\'\i}a}},
  \citenamefont {{Gaulme}}, \citenamefont {{Hall}}, \citenamefont {{Handberg}},
  \citenamefont {{Hon}}, \citenamefont {{Kallinger}}, \citenamefont {{Kang}},
  \citenamefont {{Lund}}, \citenamefont {{Mathur}}, \citenamefont {{Mints}},
  \citenamefont {{Mosser}}, \citenamefont {{{\c{C}}elik Orhan}}, \citenamefont
  {{Rodrigues}}, \citenamefont {{Vrard}}, \citenamefont {{Y{\i}ld{\i}z}},
  \citenamefont {{Zinn}}, \citenamefont {{{\"O}rtel}}, \citenamefont {{Beck}},
  \citenamefont {{Bell}}, \citenamefont {{Guo}}, \citenamefont {{Jiang}},
  \citenamefont {{Kuszlewicz}}, \citenamefont {{Kuehn}}, \citenamefont {{Li}},
  \citenamefont {{Lundkvist}}, \citenamefont {{Pinsonneault}}, \citenamefont
  {{Tayar}}, \citenamefont {{Cunha}}, \citenamefont {{Hekker}}, \citenamefont
  {{Huber}}, \citenamefont {{Miglio}}, \citenamefont {{F.~G. Monteiro}},
  \citenamefont {{Slumstrup}}, \citenamefont {{Winther}}, \citenamefont
  {{Angelou}}, \citenamefont {{Benomar}}, \citenamefont {{B{\'o}di}},
  \citenamefont {{De Moura}}, \citenamefont {{Deheuvels}}, \citenamefont
  {{Derekas}}, \citenamefont {{Di Mauro}}, \citenamefont {{Dupret}},
  \citenamefont {{Jim{\'e}nez}}, \citenamefont {{Lebreton}}, \citenamefont
  {{Matthews}}, \citenamefont {{Nardetto}}, \citenamefont {{do Nascimento}},
  \citenamefont {{Pereira}}, \citenamefont {{Rodr{\'\i}guez D{\'\i}az}},
  \citenamefont {{Serenelli}}, \citenamefont {{Spitoni}}, \citenamefont
  {{Stonkut{\.{e}}}}, \citenamefont {{Su{\'a}rez}}, \citenamefont
  {{Szab{\'o}}}, \citenamefont {{Van Eylen}}, \citenamefont {{Ventura}},
  \citenamefont {{Verma}}, \citenamefont {{Weiss}}, \citenamefont {{Wu}},
  \citenamefont {{Barclay}}, \citenamefont {{Christensen-Dalsgaard}},
  \citenamefont {{Jenkins}}, \citenamefont {{Kjeldsen}}, \citenamefont
  {{Ricker}}, \citenamefont {{Seager}},\ and\ \citenamefont
  {{Vanderspek}}}]{2020ApJ...889L..34S}%
  \BibitemOpen
  \bibfield  {author} {\bibinfo {author} {\bibfnamefont {V.}~\bibnamefont
  {{Silva Aguirre}}}, \bibinfo {author} {\bibfnamefont {D.}~\bibnamefont
  {{Stello}}}, \bibinfo {author} {\bibfnamefont {A.}~\bibnamefont
  {{Stokholm}}}, \bibinfo {author} {\bibfnamefont {J.~R.}\ \bibnamefont
  {{Mosumgaard}}}, \bibinfo {author} {\bibfnamefont {W.~H.}\ \bibnamefont
  {{Ball}}}, \bibinfo {author} {\bibfnamefont {S.}~\bibnamefont {{Basu}}},
  \bibinfo {author} {\bibfnamefont {D.}~\bibnamefont {{Bossini}}}, \bibinfo
  {author} {\bibfnamefont {L.}~\bibnamefont {{Bugnet}}}, \bibinfo {author}
  {\bibfnamefont {D.}~\bibnamefont {{Buzasi}}}, \bibinfo {author}
  {\bibfnamefont {T.~L.}\ \bibnamefont {{Campante}}}, \bibinfo {author}
  {\bibfnamefont {L.}~\bibnamefont {{Carboneau}}}, \bibinfo {author}
  {\bibfnamefont {W.~J.}\ \bibnamefont {{Chaplin}}}, \bibinfo {author}
  {\bibfnamefont {E.}~\bibnamefont {{Corsaro}}}, \bibinfo {author}
  {\bibfnamefont {G.~R.}\ \bibnamefont {{Davies}}}, \bibinfo {author}
  {\bibfnamefont {Y.}~\bibnamefont {{Elsworth}}}, \bibinfo {author}
  {\bibfnamefont {R.~A.}\ \bibnamefont {{Garc{\'\i}a}}}, \bibinfo {author}
  {\bibfnamefont {P.}~\bibnamefont {{Gaulme}}}, \bibinfo {author}
  {\bibfnamefont {O.~J.}\ \bibnamefont {{Hall}}}, \bibinfo {author}
  {\bibfnamefont {R.}~\bibnamefont {{Handberg}}}, \bibinfo {author}
  {\bibfnamefont {M.}~\bibnamefont {{Hon}}}, \bibinfo {author} {\bibfnamefont
  {T.}~\bibnamefont {{Kallinger}}}, \bibinfo {author} {\bibfnamefont
  {L.}~\bibnamefont {{Kang}}}, \bibinfo {author} {\bibfnamefont {M.~N.}\
  \bibnamefont {{Lund}}}, \bibinfo {author} {\bibfnamefont {S.}~\bibnamefont
  {{Mathur}}}, \bibinfo {author} {\bibfnamefont {A.}~\bibnamefont {{Mints}}},
  \bibinfo {author} {\bibfnamefont {B.}~\bibnamefont {{Mosser}}}, \bibinfo
  {author} {\bibfnamefont {Z.}~\bibnamefont {{{\c{C}}elik Orhan}}}, \bibinfo
  {author} {\bibfnamefont {T.~S.}\ \bibnamefont {{Rodrigues}}}, \bibinfo
  {author} {\bibfnamefont {M.}~\bibnamefont {{Vrard}}}, \bibinfo {author}
  {\bibfnamefont {M.}~\bibnamefont {{Y{\i}ld{\i}z}}}, \bibinfo {author}
  {\bibfnamefont {J.~C.}\ \bibnamefont {{Zinn}}}, \bibinfo {author}
  {\bibfnamefont {S.}~\bibnamefont {{{\"O}rtel}}}, \bibinfo {author}
  {\bibfnamefont {P.~G.}\ \bibnamefont {{Beck}}}, \bibinfo {author}
  {\bibfnamefont {K.~J.}\ \bibnamefont {{Bell}}}, \bibinfo {author}
  {\bibfnamefont {Z.}~\bibnamefont {{Guo}}}, \bibinfo {author} {\bibfnamefont
  {C.}~\bibnamefont {{Jiang}}}, \bibinfo {author} {\bibfnamefont {J.~S.}\
  \bibnamefont {{Kuszlewicz}}}, \bibinfo {author} {\bibfnamefont {C.~A.}\
  \bibnamefont {{Kuehn}}}, \bibinfo {author} {\bibfnamefont {T.}~\bibnamefont
  {{Li}}}, \bibinfo {author} {\bibfnamefont {M.~S.}\ \bibnamefont
  {{Lundkvist}}}, \bibinfo {author} {\bibfnamefont {M.}~\bibnamefont
  {{Pinsonneault}}}, \bibinfo {author} {\bibfnamefont {J.}~\bibnamefont
  {{Tayar}}}, \bibinfo {author} {\bibfnamefont {M.~S.}\ \bibnamefont
  {{Cunha}}}, \bibinfo {author} {\bibfnamefont {S.}~\bibnamefont {{Hekker}}},
  \bibinfo {author} {\bibfnamefont {D.}~\bibnamefont {{Huber}}}, \bibinfo
  {author} {\bibfnamefont {A.}~\bibnamefont {{Miglio}}}, \bibinfo {author}
  {\bibfnamefont {M.~J.~P.}\ \bibnamefont {{F.~G. Monteiro}}}, \bibinfo
  {author} {\bibfnamefont {D.}~\bibnamefont {{Slumstrup}}}, \bibinfo {author}
  {\bibfnamefont {M.~L.}\ \bibnamefont {{Winther}}}, \bibinfo {author}
  {\bibfnamefont {G.}~\bibnamefont {{Angelou}}}, \bibinfo {author}
  {\bibfnamefont {O.}~\bibnamefont {{Benomar}}}, \bibinfo {author}
  {\bibfnamefont {A.}~\bibnamefont {{B{\'o}di}}}, \bibinfo {author}
  {\bibfnamefont {B.~L.}\ \bibnamefont {{De Moura}}}, \bibinfo {author}
  {\bibfnamefont {S.}~\bibnamefont {{Deheuvels}}}, \bibinfo {author}
  {\bibfnamefont {A.}~\bibnamefont {{Derekas}}}, \bibinfo {author}
  {\bibfnamefont {M.~P.}\ \bibnamefont {{Di Mauro}}}, \bibinfo {author}
  {\bibfnamefont {M.-A.}\ \bibnamefont {{Dupret}}}, \bibinfo {author}
  {\bibfnamefont {A.}~\bibnamefont {{Jim{\'e}nez}}}, \bibinfo {author}
  {\bibfnamefont {Y.}~\bibnamefont {{Lebreton}}}, \bibinfo {author}
  {\bibfnamefont {J.}~\bibnamefont {{Matthews}}}, \bibinfo {author}
  {\bibfnamefont {N.}~\bibnamefont {{Nardetto}}}, \bibinfo {author}
  {\bibfnamefont {J.}~\bibnamefont {{do Nascimento}}, \bibfnamefont {Jose~D.}},
  \bibinfo {author} {\bibfnamefont {F.}~\bibnamefont {{Pereira}}}, \bibinfo
  {author} {\bibfnamefont {L.~F.}\ \bibnamefont {{Rodr{\'\i}guez D{\'\i}az}}},
  \bibinfo {author} {\bibfnamefont {A.~M.}\ \bibnamefont {{Serenelli}}},
  \bibinfo {author} {\bibfnamefont {E.}~\bibnamefont {{Spitoni}}}, \bibinfo
  {author} {\bibfnamefont {E.}~\bibnamefont {{Stonkut{\.{e}}}}}, \bibinfo
  {author} {\bibfnamefont {J.~C.}\ \bibnamefont {{Su{\'a}rez}}}, \bibinfo
  {author} {\bibfnamefont {R.}~\bibnamefont {{Szab{\'o}}}}, \bibinfo {author}
  {\bibfnamefont {V.}~\bibnamefont {{Van Eylen}}}, \bibinfo {author}
  {\bibfnamefont {R.}~\bibnamefont {{Ventura}}}, \bibinfo {author}
  {\bibfnamefont {K.}~\bibnamefont {{Verma}}}, \bibinfo {author} {\bibfnamefont
  {A.}~\bibnamefont {{Weiss}}}, \bibinfo {author} {\bibfnamefont
  {T.}~\bibnamefont {{Wu}}}, \bibinfo {author} {\bibfnamefont {T.}~\bibnamefont
  {{Barclay}}}, \bibinfo {author} {\bibfnamefont {J.}~\bibnamefont
  {{Christensen-Dalsgaard}}}, \bibinfo {author} {\bibfnamefont {J.~M.}\
  \bibnamefont {{Jenkins}}}, \bibinfo {author} {\bibfnamefont {H.}~\bibnamefont
  {{Kjeldsen}}}, \bibinfo {author} {\bibfnamefont {G.~R.}\ \bibnamefont
  {{Ricker}}}, \bibinfo {author} {\bibfnamefont {S.}~\bibnamefont {{Seager}}},\
  and\ \bibinfo {author} {\bibfnamefont {R.}~\bibnamefont {{Vanderspek}}},\
  }\bibfield  {title} {\bibinfo {title} {{Detection and Characterization of
  Oscillating Red Giants: First Results from the TESS Satellite}},\ }\href
  {https://doi.org/10.3847/2041-8213/ab6443} {\bibfield  {journal} {\bibinfo
  {journal} {\apjl}\ }\textbf {\bibinfo {volume} {889}},\ \bibinfo {eid} {L34}
  (\bibinfo {year} {2020})},\ \Eprint {https://arxiv.org/abs/1912.07604}
  {arXiv:1912.07604 [astro-ph.SR]} \BibitemShut {NoStop}%
\bibitem [{\citenamefont {{Thorne}}\ and\ \citenamefont
  {{Campolattaro}}(1967)}]{1967ApJ...149..591T}%
  \BibitemOpen
  \bibfield  {author} {\bibinfo {author} {\bibfnamefont {K.~S.}\ \bibnamefont
  {{Thorne}}}\ and\ \bibinfo {author} {\bibfnamefont {A.}~\bibnamefont
  {{Campolattaro}}},\ }\href {https://doi.org/10.1086/149288} {\bibinfo {title}
  {{Non-Radial Pulsation of General-Relativistic Stellar Models. I. Analytic
  Analysis for L >= 2}}},\ \bibinfo {howpublished} {Astrophysical Journal, vol.
  149, p.591} (\bibinfo {year} {1967})\BibitemShut {NoStop}%
\bibitem [{\citenamefont {{Chandrasekhar}}\ and\ \citenamefont
  {{Ferrari}}(1991)}]{1991RSPSA.432..247C}%
  \BibitemOpen
  \bibfield  {author} {\bibinfo {author} {\bibfnamefont {S.}~\bibnamefont
  {{Chandrasekhar}}}\ and\ \bibinfo {author} {\bibfnamefont {V.}~\bibnamefont
  {{Ferrari}}},\ }\bibfield  {title} {\bibinfo {title} {{On the non-radial
  oscillations of a star}},\ }\href {https://doi.org/10.1098/rspa.1991.0016}
  {\bibfield  {journal} {\bibinfo  {journal} {Proceedings of the Royal Society
  of London Series A}\ }\textbf {\bibinfo {volume} {432}},\ \bibinfo {pages}
  {247} (\bibinfo {year} {1991})}\BibitemShut {NoStop}%
\bibitem [{\citenamefont {Schutz}(2008)}]{B_F_Schutz_2008}%
  \BibitemOpen
  \bibfield  {author} {\bibinfo {author} {\bibfnamefont {B.~F.}\ \bibnamefont
  {Schutz}},\ }\bibfield  {title} {\bibinfo {title} {Asteroseismology of
  neutron stars and black holes},\ }\href
  {https://doi.org/10.1088/1742-6596/118/1/012005} {\bibfield  {journal}
  {\bibinfo  {journal} {Journal of Physics: Conference Series}\ }\textbf
  {\bibinfo {volume} {118}},\ \bibinfo {pages} {012005} (\bibinfo {year}
  {2008})}\BibitemShut {NoStop}%
\bibitem [{\citenamefont {Kokkotas}\ and\ \citenamefont
  {Schmidt}(1999)}]{Kokkotas:1999bd}%
  \BibitemOpen
  \bibfield  {author} {\bibinfo {author} {\bibfnamefont {K.~D.}\ \bibnamefont
  {Kokkotas}}\ and\ \bibinfo {author} {\bibfnamefont {B.~G.}\ \bibnamefont
  {Schmidt}},\ }\bibfield  {title} {\bibinfo {title} {{Quasinormal modes of
  stars and black holes}},\ }\href {https://doi.org/10.12942/lrr-1999-2}
  {\bibfield  {journal} {\bibinfo  {journal} {Living Rev. Rel.}\ }\textbf
  {\bibinfo {volume} {2}},\ \bibinfo {pages} {2} (\bibinfo {year} {1999})},\
  \Eprint {https://arxiv.org/abs/gr-qc/9909058} {arXiv:gr-qc/9909058}
  \BibitemShut {NoStop}%
\bibitem [{\citenamefont {Sakstein}\ and\ \citenamefont
  {Saltas}(2023)}]{sakstein_jeremy_2023_7894030}%
  \BibitemOpen
  \bibfield  {author} {\bibinfo {author} {\bibfnamefont {J.}~\bibnamefont
  {Sakstein}}\ and\ \bibinfo {author} {\bibfnamefont {I.}~\bibnamefont
  {Saltas}},\ }\bibfield  {title} {\bibinfo {title} {Dark matter-induced
  stellar oscillations},\ }\href {https://doi.org/10.5281/zenodo.7894030}
  {10.5281/zenodo.7894030} (\bibinfo {year} {2023})\BibitemShut {NoStop}%
\end{thebibliography}%



\end{document}